\documentclass[aip,jcp,amsmath,amssymb,floatfix,reprint,citeautoscript,articletitle=false,noeprint]{revtex4-1}

\usepackage[english]{babel}
\usepackage{color}
\usepackage[pdftex]{graphicx}
\usepackage[caption=false]{subfig} %
\usepackage{amsmath,amssymb,bm}
\usepackage[version=3]{mhchem}
\usepackage{verbatim}
\usepackage{multirow}
\usepackage{dcolumn}
\usepackage{float}
\usepackage{nicefrac}
\usepackage{siunitx}
\DeclareSIUnit[number-unit-product = {\,}]{\cal}{cal}
\DeclareSIUnit[number-unit-product = {\,}]{\kcal}{\kilo\cal}
\DeclareSIUnit[number-unit-product = {\,}]{\atomicunit}{a.u.}
\usepackage[english]{babel}
\usepackage[colorlinks,allcolors=black,citecolor=blue,urlcolor=blue]{hyperref}

\newcommand{\hydronium}{\cf{H3O+}}
\newcommand{\zundelp}{\cf{H5O2+}}

\begin{document}
   \title{
    High-dimensional neural network 
    potentials for solvation: The case of 
    protonated water clusters in 
    helium}
\author{Christoph Schran}
\affiliation{Lehrstuhl f\"ur Theoretische Chemie,
  Ruhr-Universit\"at Bochum, 44780 Bochum, Germany}
\author{Felix Uhl}
\affiliation{Lehrstuhl f\"ur Theoretische Chemie,
  Ruhr-Universit\"at Bochum, 44780 Bochum, Germany}
\author{J\"org Behler}
\affiliation{Universit\"at G\"ottingen,
    Institut f\"ur Physikalische Chemie,
    Theoretische Chemie,
    Tammannstr. 6, 37077 G\"ottingen, Germany}
\affiliation{Lehrstuhl f\"ur Theoretische Chemie,
  Ruhr-Universit\"at Bochum, 44780 Bochum, Germany}
\author{Dominik Marx}
\affiliation{Lehrstuhl f\"ur Theoretische Chemie,
  Ruhr-Universit\"at Bochum, 44780 Bochum, Germany}

\date{\today}

\begin{abstract}
The design of accurate
helium-solute interaction 
potentials for the simulation of 
chemically complex
molecules solvated in superfluid
helium 
has long been a cumbersome task due to the rather weak but
strongly anisotropic nature of the interactions.
We show that this challenge can be met by using a
combination of an effective pair potential for the He-He
interactions and a flexible high-dimensional neural network potential 
(NNP)
for describing the complex interaction
between helium and the solute in a pairwise additive manner.
This approach yields an excellent agreement with a mean
absolute deviation as small as \SI{0.04}{\kilo\joule\per\mol} for
the interaction energy between helium and
both, 
hydronium and Zundel cations
compared to CCSD(T) reference calculations
with
an energetically converged basis set.
The construction and improvement of the potential can be performed in a
highly
automated
way, which opens the door for applications 
to a variety of reactive molecules to study the
effect of solvation on the solute as well as the 
solute-induced structuring of the solvent.
Furthermore, we show that this NNP approach yields very convincing 
agreement with the CCSD(T) reference for properties like 
many-body
spatial 
and radial distribution functions.
This holds
for the microsolvation of 
the protonated water monomer and dimer
by a few helium atoms up to 
their solvation in bulk helium as obtained from path integral simulations at about 1~K. 
\end{abstract}

\pacs{%
}


\maketitle

\section{Introduction}
\label{sec:intro}

In recent years, helium nano droplet isolation
spectroscopy~\cite{Toennies1998/10.1146/annurev.physchem.49.1.1,
Toennies2004/10.1002/anie.200300611,
Toennies2013/10.1080/00268976.2013.802039}
has proven to be a powerful method to study 
small molecular systems 
in superfluid $^4$He environments which provide ultracold solvation conditions. 
The unique properties of this quantum fluid at temperatures in the order 
of one Kelvin allow one to examine solutes by means of
high-resolution rotational and vibrational spectroscopy
as first demonstrated in 1992.~\cite{Goyal1992/10.1103/PhysRevLett.69.933}
Although such low temperatures as well as the 
chemically
inert helium environment
circumvent many problems related to other experimental methods,
the quantum nature of the particles is pervasive at these conditions
and can not be neglected.
In such situations quantum simulation techniques employing path integrals (PI) can 
provide reliable insight by taking 
the nuclear quantum effects
explicitly into account.~\cite{Ceperley1995/10.1103/RevModPhys.67.279}
Pioneering PI simulations were able to predict the
superfluid behavior of small $^4$He clusters
\cite{Sindzingre1989/10.1103/PhysRevLett.63.1601}
which has been confirmed experimentally ten years 
later.~\cite{Grebenev1998/10.1126/science.279.5359.2083}
Since then a 
wealth
of mostly Monte Carlo (MC) based
studies have been performed for pure helium clusters,
but also for impurities in superfluid helium
or for para-H$_2$ clusters. 
Most of these studies utilize effective pair potentials for the 
helium-helium interaction, which have the advantage of being able
to reproduce the experimental observables
of pure helium under a variety of conditions with very high
accuracy.~\cite{Aziz1979/10.1063/1.438007}
In a similar spirit, also He-solute interactions are often accounted for via pairwise
additive interaction potentials.~\cite{Szalewicz2008/10.1080/01442350801933485}
This can be considered as a many-body expansion
truncated after the two-body term, which is usually acceptable
since three-body terms contribute only 
very little as has been demonstrated by computing the three-body contributions 
explicitly using extrapolated CCSD(T) calculations.~\cite{Uhl2017/10.1039/C7CP00652G}
Only recently these simulation methods have been extended to also treat reactive solute
species by coupling bosonic PI Monte Carlo (PIMC) simulations of helium 
to {\em ab initio} PI molecular dynamics (AI-PIMD) simulations of the 
solute.~\cite{Walewski2014/10.1016/j.cpc.2013.12.011,Walewski2014/10.1063/1.4870595}

A key component for all simulations of impurities in superfluid helium is 
to use a
pairwise additive interaction potential between the solute and helium
to be computationally efficient. 
The development of such potentials, however, is a rather delicate
task due to the very weak interaction between helium atoms
in the order of \SI{0.1}{\kilo\joule\per\mol}
(see e.g. Ref.~\citenum{Aziz1979/10.1063/1.438007}) 
which is therefore very sensitive to small errors.
To date, most of these 
helium-solute
potentials have been specifically designed for each
individual solute molecule using complicated, physically motivated functions
including e.g. induction and polarization terms.~\cite{
Szalewicz2008/10.1080/01442350801933485,Boese2011/10.1039/C1CP20991D,Uhl2017/10.1039/C7CP00652G}
The process of developing such potentials can therefore often become
cumbersome and even more time consuming than the final simulations 
of the molecule solvated by helium.

Here, we show how machine learning techniques
can be used to both, simplify and 
automate
the task of representing the pair interaction
energy between chemically complex solute molecules and helium.
For this purpose we use the high-dimensional neural network (NN) methodology
of Behler and Parrinello~\cite{Behler2007/10.1103/PhysRevLett.98.146401,
Behler2011/10.1039/C1CP21668F,
Behler2014/10.1088/0953-8984/26/18/183001,
Behler2015/10.1002/qua.24890,
Behler2016/10.1063/1.4966192,
Behler2017/10.1002/anie.201703114}
that has already been proven to be well suited for the description 
of potential energy surfaces of a variety of 
complex molecular 
systems including solvents~\cite{Morawietz2016/10.1073/pnas.1602375113,
Bingqing2016/10.1021/acs.jpclett.6b00729,
Kapil2016/10.1063/1.4971438},
solvated ions~\cite{Hellstrom2017/10.1039/C6CP06547C,
Hellstrom2017/10.1021/acs.jpcb.7b01490}
and surfaces~\cite{Natarajan2016/10.1039/C6CP05711J,
Quaranta2017/10.1021/acs.jpclett.7b00358}
(for a comprehensive list see e.g. Ref.~\citenum{Behler2017/10.1002/anie.201703114}
and Ref.~\citenum{Behler2016/10.1063/1.4966192}
where also references to alternative modern techniques are provided). 

Due to the intrinsic flexibility of NNs, it is possible
to easily identify atomic configurations needed for an improvement of the potential~\cite{Behler2014/10.1088/0953-8984/26/18/183001},
which substantially reduces the number of required reference calculations 
compared to traditional approaches that utilize physically motivated
functions.~\cite{Szalewicz2008/10.1080/01442350801933485,
Boese2011/10.1039/C1CP20991D,Uhl2017/10.1039/C7CP00652G}
Evidently, this property greatly reduces the computational cost due to 
reference calculations, which might be the decisive factor if
correlated and thus demanding quantum chemistry methods such as CCSD(T)
shall be used. 
Moreover, it
enables the automation of the 
development of interaction potentials
such that the simulation of a variety of reactive molecules and complexes
solvated by superfluid helium becomes feasible. 
In this first study along these lines, we focus
on the development of two He-solute NN potentials (NNPs) for the
description of 
hydronium (\hydronium{}) and  Zundel (\zundelp{}) cations, respectively, as the
two smallest protonated water clusters in the sequence
of stepwise solvation of a proton by water molecules in order to 
shed new light on proton transfer processes
in our future work. 

The outline of the paper is as follows: 
We first describe the relevant aspects of the NN methodology and 
the computational details in Sec.~\ref{sec:methods}.
The fitting procedure of the NNPs including
the approach to identify relevant training configurations for
improvements of the potentials are presented subsequently in
Sec.~\ref{sec:res_nnfit}.
This is essentially a three step procedure
progressing from simple sampling to importance sampling 
strategies to generate the training data
in which
we first obtain configurations of the solute molecules
as the basis for the generation of He-solute pairs.
In a second step, the properties of the NNs are used to iteratively
target He-solute pairs needed for an improvement
of the interaction potential, thus providing 
a powerful tool to reduce
the number of expensive reference calculations.
Afterwards, the NNPs are applied in 
full path integral quantum
simulations including many solvent He atoms to 
generate, detect and
incorporate those solvation 
structures
that are not already covered by the included He-solute pairs.
Finally, the converged pairwise additive He-solute NNPs are applied to
simulate helium around selected fixed solute configurations
starting with microsolvation up to the helium bulk phase, see 
Sec.~\ref{sec:micro} and~\ref{sec:bulk}, in order to
assess 
the quality of the developed potentials,
followed by conclusions and outlook.

\section{Methods}
\label{sec:methods}
\subsection{Determination of the Training and Test Configurations}
\label{sec:refval}

To generate reference structures of the solute molecules similar to the 
configurations in superfluid helium, we performed density functional theory
based AI-PIMD simulations~\cite{Marx2009} of the bare
solute cations \hydronium{} and \zundelp{} in vacuum
but at the same ultralow temperature.
This approach is justified
since it is known that He-solvation has only a minor effect on
solute structure.~\cite{Toennies2013/10.1080/00268976.2013.802039}
These simulations have been carried out with our in-house
developer's version of the 
\texttt{CP2k} program package~\cite{CP2K,Hutter2014/10.1002/wcms.1159} 
in a 9.0~\AA{} cubic box with cluster 
(i.e. non-periodic)
boundary conditions.
The electronic structure in the AI-PIMD simulations was solved on-the-fly using the
Quickstep module~\cite{VandeVondele2005/10.1016/j.cpc.2004.12.014}
by applying the RPBE exchange correlation functional~\cite{Hammer1999/10.1103/PhysRevB.59.7413} together with
the D3 dispersion correction~\cite{Grimme2010/10.1063/1.3382344} up to two-body terms.
The charge density was represented on a grid up to a plane wave cutoff of 500~Ry.
The TZV2P basis set together with Goedecker-Teter-Hutter
pseudopotentials to replace the
core electrons in the oxygen atoms~\cite{Goedecker1996/10.1103/PhysRevB.54.1703}
was used for the description of the Kohn-Sham orbitals.
The SCF cycles were converged to an error of $\epsilon_\text{SCF} = 10^{-7}~\text{Ha}$.
This electronic structure setup is known to reproduce
many properties of water in good agreement with 
experiment.~\cite{Morawietz2013/10.1021/jp401225b,
Forster-Tonigold2014/10.1063/1.4892400,
Morawietz2016/10.1073/pnas.1602375113}
For each solute molecule \SI{27.5}{\pico\second} AI-PIMD trajectories
were generated with a time step of \SI{0.25}{\femto\second}
at a temperature of \SI{1.67}{\kelvin}
where the path integral has been discretized in terms of 48 replicas.
The PIGLET algorithm~\cite{Ceriotti2012/10.1103/PhysRevLett.109.100604}
recently extended to ultralow temperatures\cite{Uhl2016/10.1063/1.4959602}
was applied to sample the canonical quantum distribution.
At the beginning of each trajectory, \SI{2.5}{\pico\second} 
have been discarded as equilibration starting from the 
optimized
minimum energy configuration.

From these reference ensembles we extracted structures
for the hydronium and Zundel cations with a time lag 
of \SI{25}{\femto\second} from all 
48
PI replicas, resulting
in \num{48000} uncorrelated
path integral quantum
configurations.
These structures
have been used as the basis for the generation of
helium sample positions as described later in detail,
thus producing a large ensemble of He-solute
pair configurations for the following reference electronic structure calculations.
Note that we restrict the interaction potentials to He-solute two-body interactions,
since higher-order many-body terms have recently been shown to be
negligible~\cite{Uhl2017/10.1039/C7CP00652G}.
The He-solute interaction energies were calculated
within the usual supermolecule approach 
via CCSD(T),
being broadly considered to be the ``gold standard'' of quantum chemistry,
employing the aug-cc-pVTZ basis
set~\cite{Dunning1992/10.1063/1.462569,
Woon1994/10.1063/1.466439} with a
counterpoise correction\cite{Boys1970/10.1080/00268977000101561} 
to
correct for the basis set superposition error,
dubbed AVTZcp in the following.
This electronic structure setup is shown to produce 
negligible remaining basis set errors compared to the
complete basis set limit as presented in 
Section~I.A of the SI.
All interaction energy calculations for the He-solute structures
were performed with the \texttt{Molpro} program package\cite{MOLPRO}.

\subsection{Neural Network Fitting}
\label{sec:modelpot}

The construction of the NNPs for the description of the helium-solute
interaction energies is achieved by following a similar
procedure as established by Behler and
coworkers.~\cite{Behler2007/10.1103/PhysRevLett.98.146401,
Behler2014/10.1088/0953-8984/26/18/183001}
However, instead of the total energy of the system
as before,
the He-solute interaction energy
is used as target for the fit.
The two-body interaction energy $E^\text{int}$
in any specific He-solute configuration is
then
calculated as in
the realm of 
conventional high-dimensional NNPs from
atomic contributions $E^\text{NN}_i$ for which independent atomic NNs are fitted,
\begin{align}
    E^\text{int} = \sum_{i}^{N_\text{atom}} E^\text{NN}_i 
\enspace , 
\end{align}
where $N_\text{atom}$ denotes the number of atoms in the system.
In order to construct the analytic structure-atomic energy relation,
first the atomic coordinates of the system are transformed
to sets of many-body atom-centered radial and angular symmetry
functions~\cite{Behler2011/10.1063/1.3553717} 
as usual,
which serve as input vectors for the atomic NNs.
This ensures the required invariances
of the energy with respect to translations and rotations of the system
as well as to permutations of atoms of the same element.
For both types of symmetry functions employed here, a cutoff function
\begin{align}
\label{eq:cut}
    f_\mathrm{c}(R_{ij})=
    \begin{cases}
        0.5\cdot \left[\cos\left(\frac{\pi R_{ij}}{R_\mathrm{c}}\right) + 1\right] \qquad &\text{for } R_{ij}\leq R_\mathrm{c}\\
        0  &\text{else}
    \end{cases}
\end{align}
depending on the distance $R_{ij}$ between the central atom $i$ and
a neighboring atom $j$ is used to define the atomic environment
up to a certain cutoff radius $R_\mathrm{c}$.
The radial arrangement of the atoms within this cutoff sphere is accounted for
by a product of a Gaussian and the cutoff function according to Eq.~(\ref{eq:cut}),
\begin{align}
    G^2_i = \sum_j e^{-\eta(R_{ij}-R_\mathrm{s})^2}\cdot f_\mathrm{c}(R_{ij}) 
\enspace , 
\end{align}
where different regions around the central atom $i$ can be probed
by adapting the width of the Gaussian $\eta$ and the shifting parameter $R_\mathrm{s}$.
To complement the description of the environment around each atom, angular functions
of the form
\begin{align}
    G^4_i = 2^{1-\zeta} \sum_{j,k\neq i,j \neq k} &(1+\lambda\cos\theta_{ijk})^{\zeta}\cdot e^{-\eta(R_{ij}^2+R_{ik}^2+R_{jk}^2)} \nonumber \\
    &\cdot f_\mathrm{c}(R_{ij})\cdot f_\mathrm{c}(R_{ik})\cdot f_\mathrm{c}(R_{jk})
\end{align}
are employed that depend on the angle $\theta_{ijk}$ between the central atom $i$ and
two neighbors $j$ and $k$, where $i,j$ and $k$ can be any atom of the He-solute complex.
Different angular regions are probed by adjusting the
exponent $\zeta$. 
The parameter $\lambda$, which can have values of $+1$ or $-1$,
is used to shift the maximum of the cosine either to $\pi$ or $2\pi$.

A set of these symmetry functions for each element, 
as specified in 
Section~I.D of
the SI, transforms
the coordinates of the system to be employed as input for the NN. 
We finally selected different sets for
the He-hydronium and He-Zundel interaction potentials
in order to reduce the number of 
symmetry
functions to a minimum.
For He-hydronium we employed \num{60} 
such
functions, while \num{79} are used for the He-Zundel NNP.
The values of each symmetry function are furthermore centered
around the average value of the training set and normalized to
values between zero and one
following the usual procedure.

These vectors serve as the input for the atomic NNs,
which consist in all cases of two hidden layers with 25 nodes each, and yield the
atomic energy contributions
that sum up 
to the total interaction energy.
Bias nodes with weight parameters $b$ were attached to all layers but the input layer.
Note that the evaluation of a
NN
node value $y$ is
essentially a linear transformation of the values of the previous layer $y_i$
by the associated weight parameters $a_i$,
that is further transformed by a non-linear activation function $f$
\begin{align}
 y = f \left(b +\sum_{\text{previous}}^{\text{layer}} y_i \cdot a_i \right)
\enspace 
.
\end{align}
We used the hyperbolic tangent in all hidden layers 
and a linear activation function for the output layer
in order to prevent a confined range of output values.

The NNPs are constructed by first splitting the
set of CCSD(T) reference data
into a training set (90\%)
and an independent test set (10\%) of He-solute configurations.
Subsequently, the weight parameters of the NNs are iteratively optimized to
minimize the error of the training set, while the test set provides an estimate 
for the transferability to structures not included
in the training set and is used to detect over fitting.
Learning was achieved by
optimizing the weights $a$ and $b$ according to the 
adaptive global extended Kalman filter~\cite{Shah1992/10.1016/S0893-6080(05)80139-X,
Blank1994/10.1002/cem.1180080605,Witkoskie2005/10.1021/ct049976i}
as implemented in our in-house program \texttt{RuNNer}~\cite{Behler/Runner}.

\subsection{Path Integral Simulations}
\label{sec:PISimulations}

In order
to
incorporate the solvation effect on the solute structure and to use 
NNPs in (ab initio) path integral simulations,
the NNs
have been implemented 
in our in-house version of the \texttt{CP2k} program package~\cite{CP2K}
to provide energies and forces 
for
the hybrid PIMD / bosonic PIMC
technique~\cite{Walewski2014/10.1016/j.cpc.2013.12.011,
Walewski2014/10.1063/1.4870595}.

To identify solute configurations outside the 
structural
range of the
vacuum ensemble formed by helium-solute pairs, which might emerge during helium solvation, 
we conducted short \SI{10}{\pico\second}
AI-PIMD/PIMC simulations~\cite{Walewski2014/10.1016/j.cpc.2013.12.011,
Walewski2014/10.1063/1.4870595} of the fully flexible solutes in bulk helium
starting from equilibrated vacuum solute configurations
and coupling the solutes to the helium environment
via the NNPs.
The solutes were simulated together with 98 and 88 helium atoms
for the \hydronium{} and \zundelp{} cation, respectively,
in a truncated octahedron cell with periodic boundary conditions
and a distance between the parallel square faces of 19.117~\AA{}.
The number of helium atoms placed in this box were chosen 
according to a comparison of radial distribution 
functions (RDFs) as presented in 
Section~I.C of 
the SI.
Helium was treated using the pair density matrix 
approximation~\cite{Ceperley1995/10.1103/RevModPhys.67.279} 
together with the Aziz He-He pair interaction potential~\cite{Aziz1979/10.1063/1.438007}.
The temperature was set to 1.67~K and the high-temperature 
pair density matrix at 80~K was used together with 48 beads
in order to discretize the low-temperature density matrix at 1.67~K.
The solute was described by the same setup as for the vacuum simulations 
presented in Sec.~\ref{sec:refval}.
Following our earlier work 
on the solvation of molecules in superfluid helium~\cite{Walewski2014/10.1016/j.cpc.2013.12.011,Walewski2014/10.1063/1.4870595},
the He-solute interactions are discretized using the primitive approximation
whereas the development of higher-order actions involving NNP couplings are a subject of future research.
In between any two AI-PIMD steps, \num{10000} PIMC steps were performed
to ensure sufficient helium sampling
according to the LaBerge-Tully ``MDMC
algorithm''.~\cite{LaBerge2000/10.1016/S0301-0104(00)00246-9}

In order to validate the quality of our two final NNPs compared to CCSD(T)/AVTZcp data,
we evaluated the interaction potential of five
frozen structures of the target solutes on a cubic grid.
Bulk helium and helium microsolvation can be simulated surrounding
the clamped solute structures by evaluating the 
solute-helium interaction on the grid using
the nearest neighbor approach, thus assigning
the interaction in continuous space to the nearest grid point.
The structures were centered in a cubic grid with 50 grid points in each dimension
and a grid spacing of 0.25~\AA{}.
This results in \num{125000} CCSD(T) calculations for each structure 
that are not used in the fitting process, but
only for the validation of the potentials.
It should be noted that this 
rigorous ``real life""
validation step requires one order of magnitude more
calculations than the actual NNP fitting procedure,
and is used here 
exclusively
to demonstrate the power of the proposed approach
as proof of concept.

Subsequent helium PIMC simulations were performed with the same
settings as for the flexible solute simulations, but
involved 20 
PIMC
walkers with randomized helium starting 
configurations and different random number seeds for bulk helium.
Each walker generated \num{1000000} structures with \num{1000} 
PIMC steps
in between to guarantee uncorrelated configurations.
Microsolvation was simulated for 1, 2, 4, 6, 10 and 14 helium atoms
with 100 
PIMC
walkers in a 15~\AA{} droplet radius
and otherwise using the same settings as for the bulk.
To estimate if the simulation length is sufficient,
we compared spatial properties for a selected configuration
obtained with the described settings to
those resulting from ten times improved statistics
as presented in Sec.~II.C of the SI.
Since these observables are essentially the same,
it can be concluded that the here reported
properties are statistically converged.
Bosonic helium exchange~\cite{Walewski2014/10.1016/j.cpc.2013.12.011,
Walewski2014/10.1063/1.4870595} was disabled for all calculations to
increase the sampling efficiency.

\section{Results and Discussion}
\label{sec:results}
\subsection{%
Automated 
Neural Network Fitting Procedure 
and Iterative Improvement}
\label{sec:res_nnfit}

In order to describe the interaction potential between
protonated water clusters and helium, we developed the following
procedure to select relevant configurations for the training of the NNP.
We started by using converged ensembles of the
solutes in vacuum obtained as described in Sec.~\ref{sec:refval}
as the basis for the generation of helium sample positions.
This is based on the assumption that helium solvation
has only a small effect on the solute structure and will
be accounted for in a subsequent step.
In contrast to interaction potentials based on physical principles, it is not sufficient to
use only stationary points on the potential energy surface
of the solutes since 
NNs
are not able to extrapolate.
Therefore, the whole range of possible configurations must be 
represented by the training set in contrast to traditional
approaches
to generate helium-solute pair potentials.~\cite{Boese2011/10.1039/C1CP20991D,Uhl2017/10.1039/C7CP00652G}
However, to restrict the reference calculations to the
relevant parts of configuration space, we used a 
simple sampling MC scheme on atom centered grids around 
the solute atoms.
Helium configurations very close to the solute molecule
are very high in energy due to Pauli repulsion.
We therefore determined a lower cutoff radius around each
atom type by analyzing a radial scan for the 
minimum energy structures of the He-H$_3$O$^+$ adduct.
We afterwards restricted our atom centered
grids to helium positions with distances exceeding 
2.05~\AA{} for oxygen and 1.25~\AA{} for hydrogen atoms.
A detailed discussion of the radial scan can be found in 
Section~I.B of
the SI.
Moreover, configurations far away from the solute feature
very small and slowly changing interaction energies and thus this region does not
need to be sampled as extensively as the region closer to
the solute.
We therefore applied Euler-Maclaurin radial
grids~\cite{Murray1993/10.1080/00268979300100651} with
$N_\text{rad}=$ 25 points which use smaller intervals close to the center.
Helium configurations farther away than 10.0~\AA{} were excluded
explicitly, since they exhibit interaction energies smaller than
\SI{0.01}{\kilo\joule\per\mole} and are usually outside the
range encountered in our envisaged helium bulk simulations.
This consideration results in the following interatomic distance constraints
\begin{align}
\label{eq:cutoff}
    r_\text{X-He} = \begin{cases}
        2.05~\text{\AA{}}  < r_\text{O-He} <  10.0~\text{\AA{}} \\
        1.25~\text{\AA{}} < r_\text{H-He} <  10.0~\text{\AA{}}
           \end{cases}
\end{align}
for possible helium configurations around the solute ensembles.

For each distance on the Euler-Maclaurin radial grid we generated
$N_\text{ang}=$ 110 additional angular points using the Lebedev 
quadrature formula~\cite{Lebedev1977/10.1007/BF00966954}
and each shell 
was
rotated around a random axis by a
random angle.
This product of radial and angular grids centered around
each solute atom of every solute configuration extracted
from the vacuum reference simulations
results in roughly 
$N_\text{sol. struc.}\cdot N_\text{sol. atoms} \cdot N_\text{rad} \cdot N_\text{ang}$
He-solute structures, some of which
are excluded due to the lower and upper cutoff
according to Eq.~(\ref{eq:cutoff}).
On this large ensemble of structures, we performed MC
simple sampling to extract He-solute configurations
for \hydronium{} and \zundelp{}.
With this procedure only 1000 
statistically independent
helium-solute 
configurations
were generated as the reference set for the first 
stage of our NNP generation protocol.

\begin{figure*}[th!]
    \includegraphics[width=1.0\textwidth]{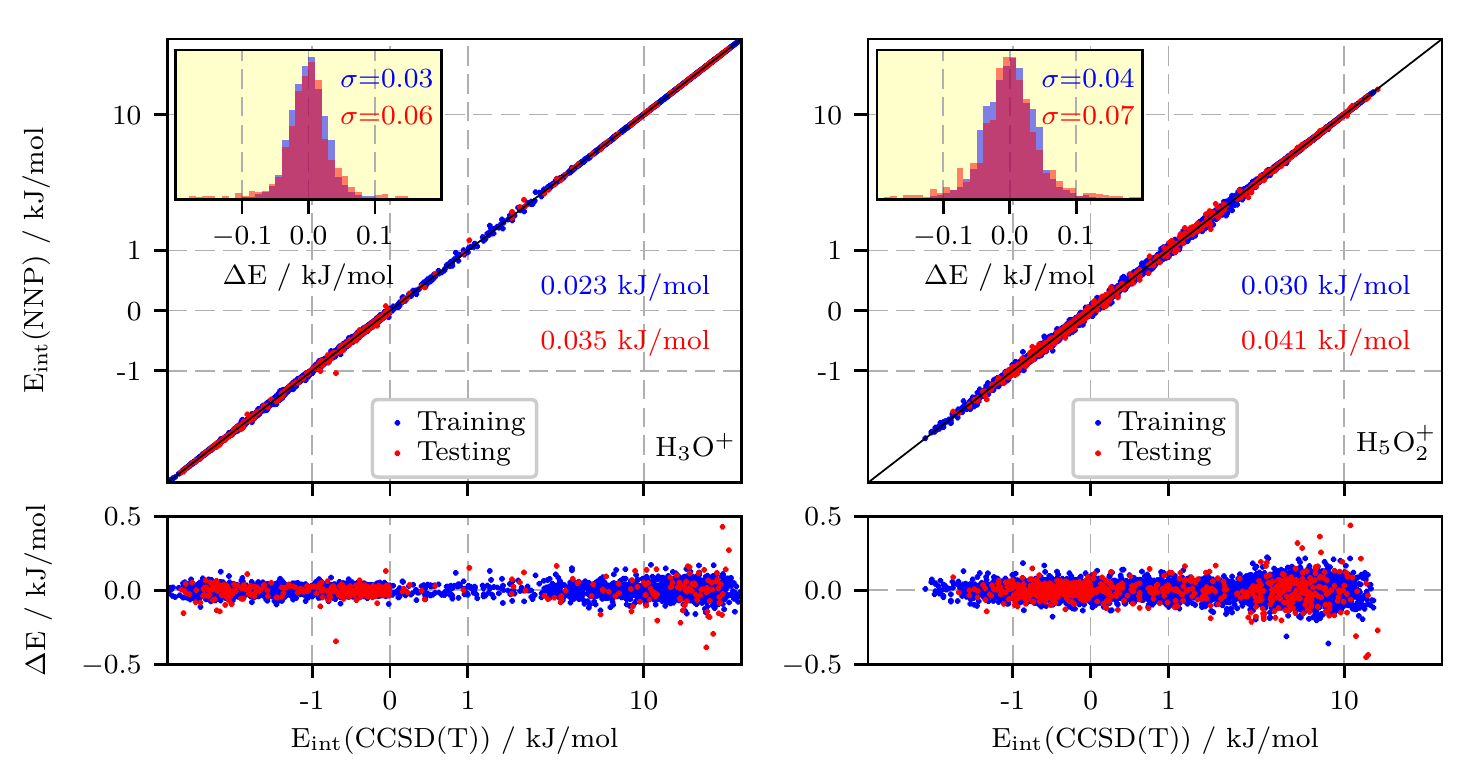}

    \caption{Correlation of the interaction energy from explicit CCSD(T)/AVTZcp 
    calculations and NNP prediction for the final reference data set, see text.
    Results for the He-hydronium (\hydronium{}) interaction potential are shown in the left panel,
    while the results of the He-Zundel (\zundelp{}) interaction potential are depicted
    in the right panel. 
    The mean absolute difference for the training and test
    set are depicted in blue and red, respectively. 
    The lower two panels show the energy
    differences between 
    coupled cluster
    reference and NNP prediction over the whole range of reference energies,
    while the insets show the histograms over the energy differences 
including 
the corresponding standard deviations $\sigma$ in the respective color
    which are well below the He-He interaction energy.
\label{fig:fit_int}}
\end{figure*}

Due to the flexibility of NNs, energy predictions
of two distinct NNPs vary considerably for
structures differing significantly from the training
configurations.~\cite{Behler2014/10.1088/0953-8984/26/18/183001}
This allows us to identify 
most efficiently
configurations that are needed for an improvement of the NNP
without performing computationally demanding 
CCSD(T) calculations for the identification of these points.
To systematically extend the reference set
we therefore developed the following strategy.
In each refining stage, additional \num{100000}
He-solute configurations for both solutes were generated
by the same procedure as before with slight modifications.
Structures in the Pauli repulsion regime and in the interaction well feature larger
absolute interaction energies than structures farther away from the solute molecule.
The resulting absolute difference between two NNPs is thus
also larger compared to interaction energies close to zero 
(assuming that the prediction is not completely unreasonable).
As before, we
randomly selected \num{1000} structures
of the solute vacuum ensemble as the basis for
helium sampling positions.
However, to circumvent a biased selection of 
structures in the interaction wells and in the Pauli repulsion dominated region,
we afterwards
randomly selected one of the following three intervals for the radial
grids around each solute atom to systematically improve different 
regions of the interaction potential, namely
\begin{align}
    r_\text{X-He}^1 &= \begin{cases}
        2.05~\text{\AA{}}  < r_\text{O-He} <  10.0~\text{\AA{}} \\
        1.25~\text{\AA{}} < r_\text{H-He} <  10.0~\text{\AA{}}
           \end{cases} \nonumber\\
    r_\text{X-He}^2 &= \begin{cases}
        2.5~\text{\AA{}}  < r_\text{O-He} <  10.0~\text{\AA{}} \\
        1.5~\text{\AA{}} < r_\text{H-He} <  10.0~\text{\AA{}}
           \end{cases}\\
    r_\text{X-He}^3 &= \begin{cases}
        4.0~\text{\AA{}}  < r_\text{O-He} <  10.0~\text{\AA{}} \\
        3.0~\text{\AA{}} < r_\text{H-He} <  10.0~\text{\AA{}}
           \end{cases}.\nonumber
\end{align}
The two additional cutoffs are chosen according to the first and
second coordination shell around the solutes.
Afterwards, we evaluated the interaction energies for the set of additional
configurations with two different NNPs 
and extracted the \num{1000} structures with the highest energy differences.
For each of these configurations we performed CCSD(T) reference calculations 
and the new data was then combined with the old CCSD(T) data set to serve as
input for the fit of the refined NNPs at the next stage. 
This procedure allows us to systematically improve the 
NNPs,
while substantially lowering the number of 
expensive CCSD(T) reference calculations.
Overall, we iteratively improved the NNP for He-hydronium in 12
such refinement stages, 
while 21 
stages
were used for the He-Zundel NNP to account for
the larger number of degrees of freedom.
We thus used 
only
\num{12000} and \num{21000} reference CCSD(T) calculations for
the hydronium and Zundel complex, respectively.

In a next step we explicitly incorporated the solvation effect of
helium on the solute structure by conducting short
AI-PIMD/PIMC 
simulations
of the two fully flexible solutes in bulk helium
coupled via the refined NNPs as described in 
detail in Sec.~\ref{sec:PISimulations}.
During these simulations, we are able to identify structures
that are missing in our training set by
comparing the encountered symmetry function values to the
range of values in the training set.
Structures outside the range are extrapolating and therefore
have to be included for a reliable 
NN
description of the interaction energy.
During the short simulations that already evaluate the NNP 
many million times, we 
identified and 
extracted \num{3545} and \num{4101}
structures that feature symmetry function values outside
the range of the training set for the \hydronium{} and 
\zundelp{} molecule, respectively, and performed reference 
CCSD(T) 
calculations for all of them. 
They were afterwards combined with the previous training sets
to parameterize the final interaction potentials.

The correlation between the reference CCSD(T)/AVTZcp energies 
and the NNP interaction potential of the final training and
test sets is shown in Fig.~\ref{fig:fit_int}.
For both solutes almost perfect correlation is achieved for the training
as well as the test set over the whole range of interaction energies.
The Zundel cation features overall smaller helium interaction energies
than the hydronium cation. 
The strongest interaction
for He-hydronium in the reference set is \SI{-5.77}{\kilo\joule\per\mole},
while for He-Zundel it is only \SI{-3.35}{\kilo\joule\per\mole}.
This is due to a larger charge localization in the smaller
\hydronium{} species
that therefore results in stronger interaction with helium.
The lower panels of the figure additionally show the
deviation of the predicted energies over the whole range of the reference set. 
These are consistently small, with slightly larger deviations 
in the repulsive regime which, however, also features
one order of magnitude larger energies.
The
mean absolute differences
(\hydronium{}: \SI{0.023}{\kilo\joule\per\mol} training set,
\SI{0.035}{\kilo\joule\per\mol} test set; 
\zundelp{}: \SI{0.030}{\kilo\joule\per\mol} training set,
\SI{0.041}{\kilo\joule\per\mol} test set)
are of comparable order in the training and test set 
as well as for both networks.
These values are considerably lower than for
other interaction potentials that have been 
obtained in a traditional fitting approach to physically 
derived He-solute interaction potentials
of other systems.~\cite{Boese2011/10.1039/C1CP20991D,
Uhl2017/10.1039/C7CP00652G}
In addition, as shown in the insets of
the upper panels of Fig.~\ref{fig:fit_int}, the
histograms over the deviations of the predicted energies
for the training and test set
are very narrow with standard deviations
in the order of \SI{0.05}{\kilo\joule\per\mole} and thus well below
the He-He interaction of \SI{0.1}{\kilo\joule\per\mole}.
It can therefore be concluded that 
NNs
are able to fit interaction energies with very high precision.
In addition, our presented method of selecting reference configurations
allows for fast improvement of the interaction potentials and
substantially reduces the number of computationally demanding CCSD(T) calculations
compared to traditional approaches~\cite{Boese2011/10.1039/C1CP20991D,
Uhl2017/10.1039/C7CP00652G}.
Finally, we stress that the protocol developed
for hydronium and Zundel cations can be readily 
automated 
and thus generalized to other solutes in helium.

\subsection{Helium Microsolvation}
\label{sec:micro}

The canonical way to evaluate the quality of NNPs
is to compare to calculations performed with the method used 
to determine the reference data set.
In our case this would imply to run our envisaged hybrid
AI-PIMD/PIMC quantum simulation and evaluate the CCSD(T) energy in
each and every AI-PIMD and PIMC step as well as the forces in every AI-PIMD step
of the simulation.
Unfortunately, this method is out of scope,
since one MC cycle requires already around
40~million (number of helium atoms $N_\text{He}$ times number
of path integral replica $N_\text{Rep}$ times number of 
auxiliary MC steps $N_\text{MC}$)
evaluations of the interaction energy.
Note that it is
of course
possible to compare radial scans
around selected solute structures as demonstrated for
the hydronium cation in 
Section~I.B of
the SI.
However, this approach examines only
a very small fraction of the full configuration space
and does not incorporate the influence of He-He
interactions.

\begin{figure}[t!]
    \begin{minipage}[t]{0.25\columnwidth}
    \includegraphics[trim=400 750 400 700,clip,width=1.0\textwidth]{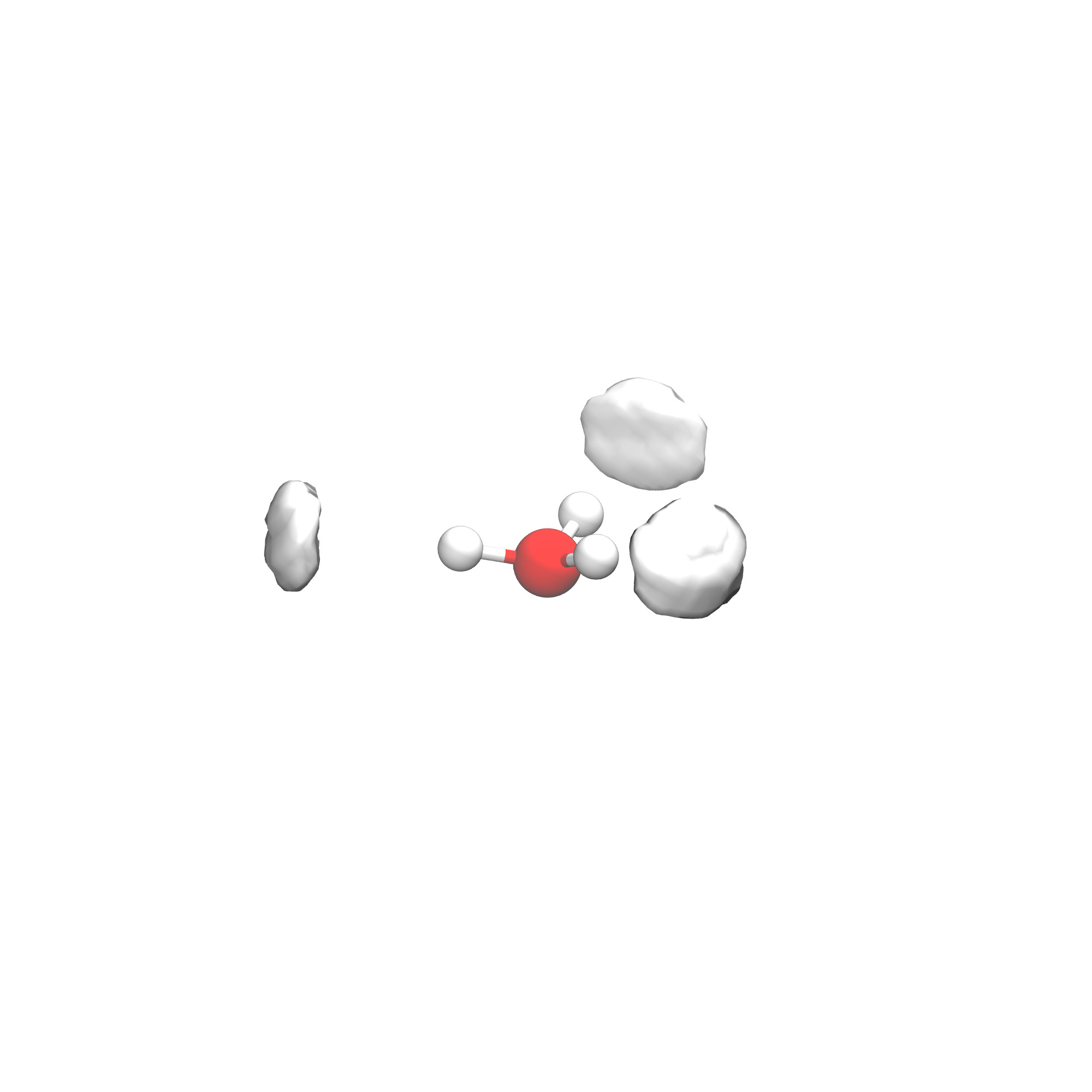}
    \end{minipage}
    \begin{minipage}[t]{0.25\columnwidth}
    \includegraphics[trim=400 750 400 700,clip,width=1.0\textwidth]{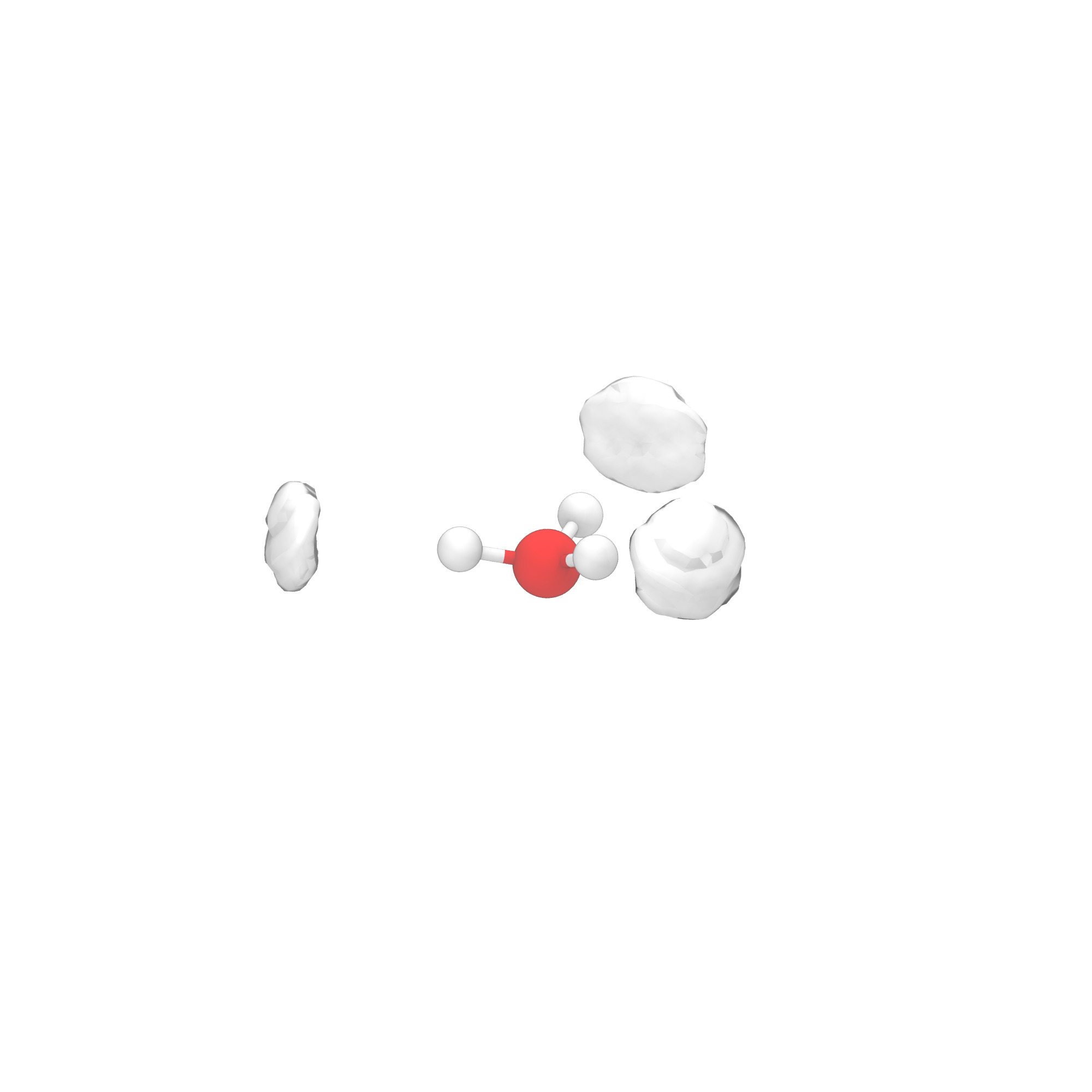}
    \end{minipage}\\
    \begin{minipage}[t]{0.25\columnwidth}
    \includegraphics[trim=400 750 400 700,clip,width=1.0\textwidth]{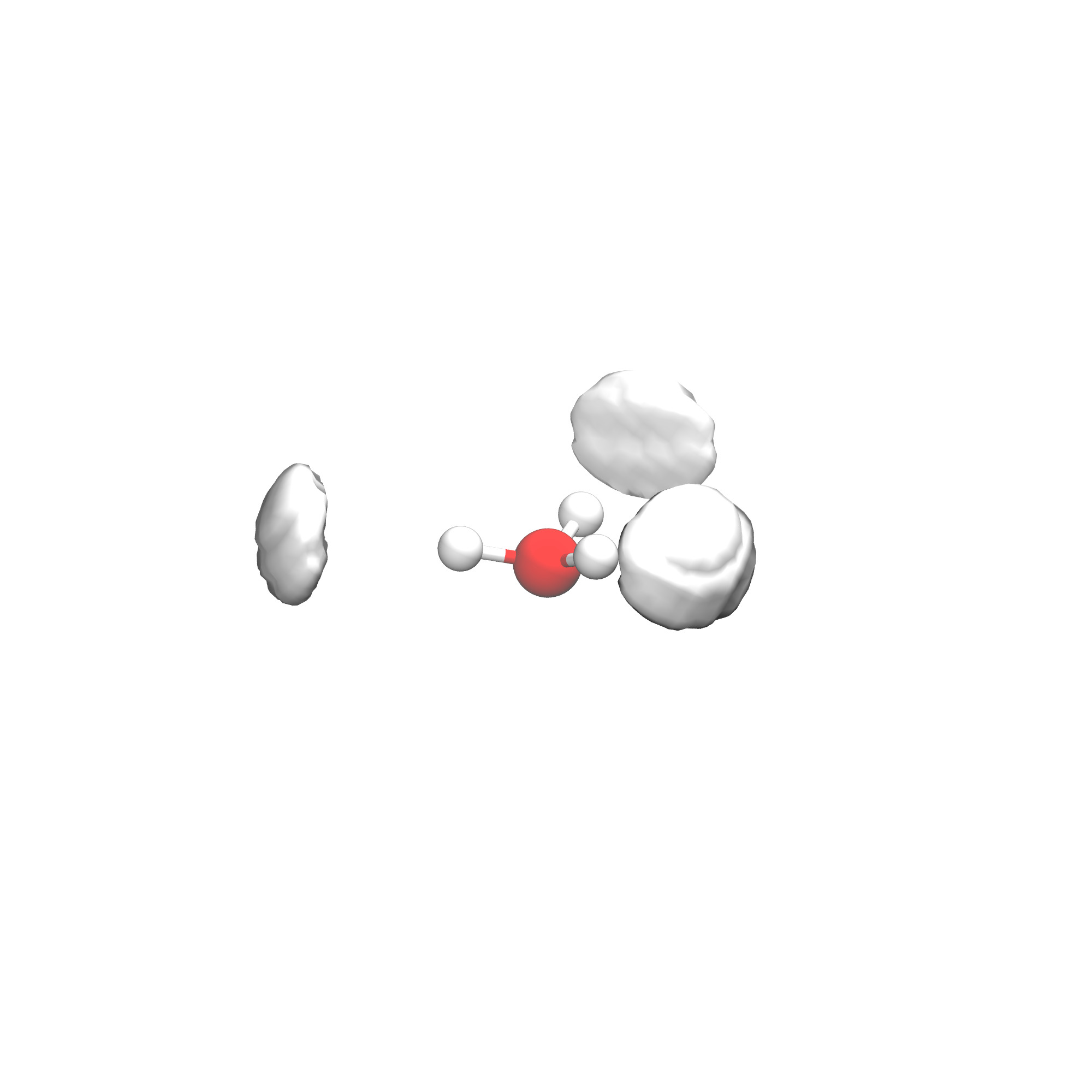}
    \end{minipage}
    \begin{minipage}[t]{0.25\columnwidth}
    \includegraphics[trim=400 750 400 700,clip,width=1.0\textwidth]{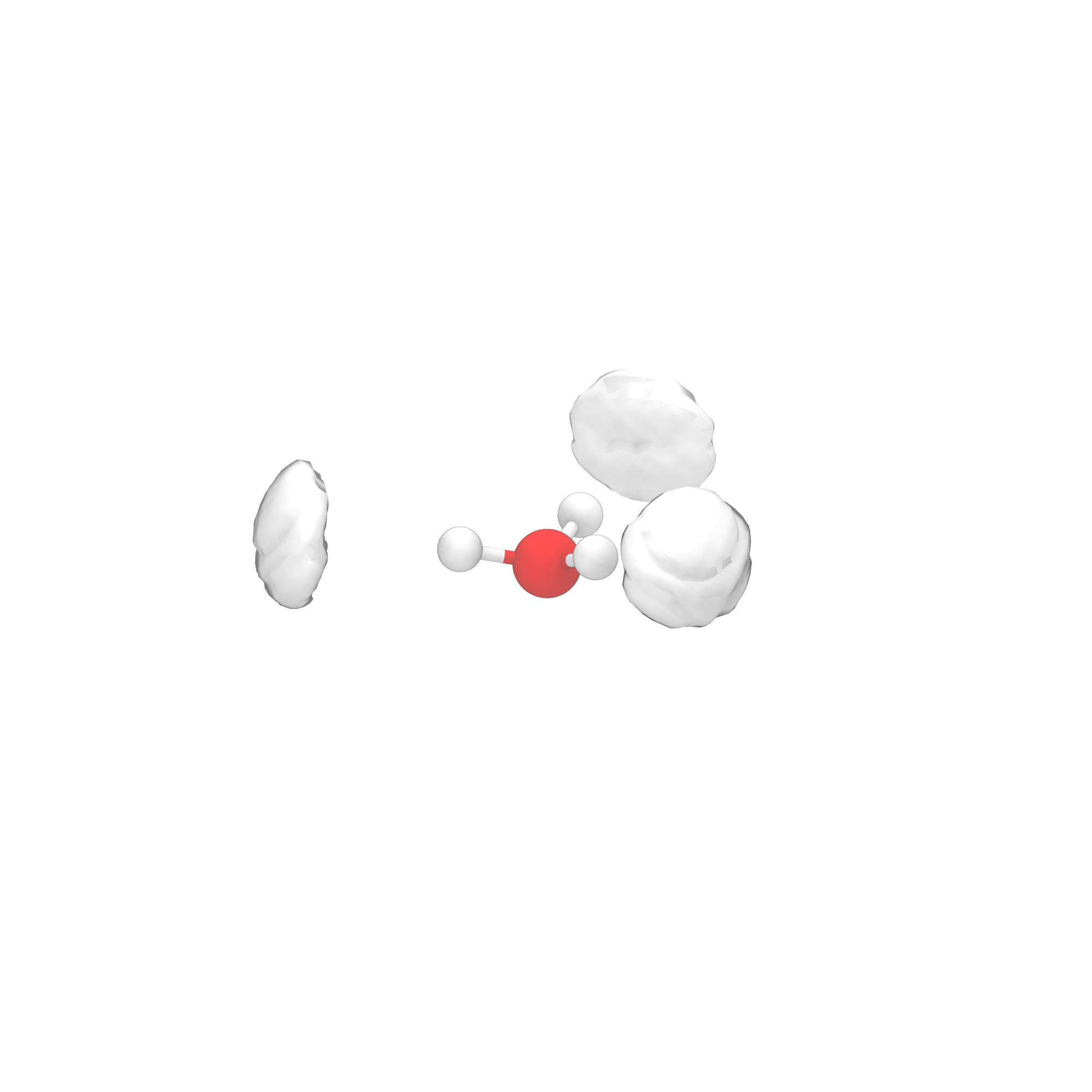}
    \end{minipage}\\
    \begin{minipage}[t]{0.25\columnwidth}
    \includegraphics[trim=400 750 400 650,clip,width=1.0\textwidth]{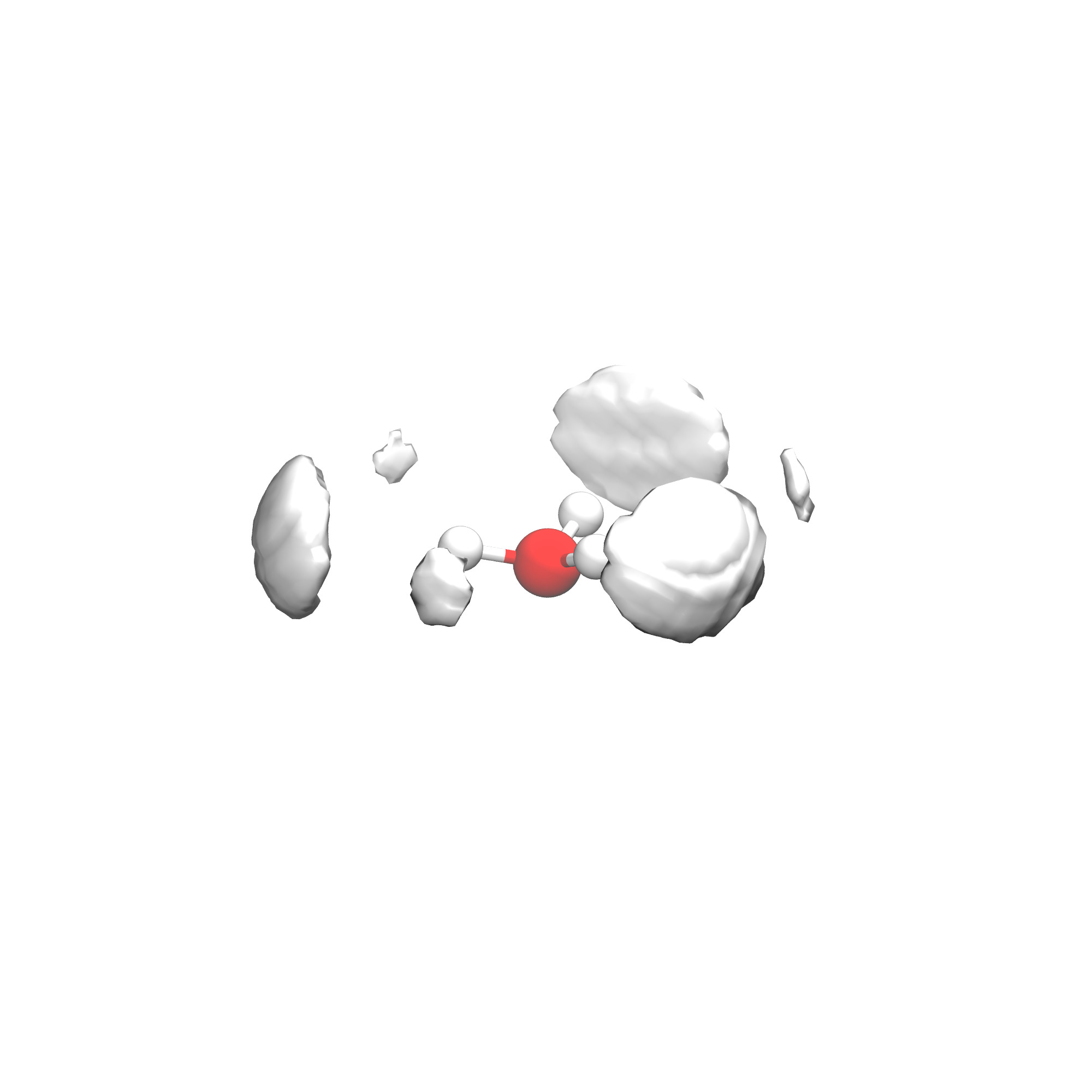}
    \end{minipage}
    \begin{minipage}[t]{0.25\columnwidth}
    \includegraphics[trim=400 750 400 650,clip,width=1.0\textwidth]{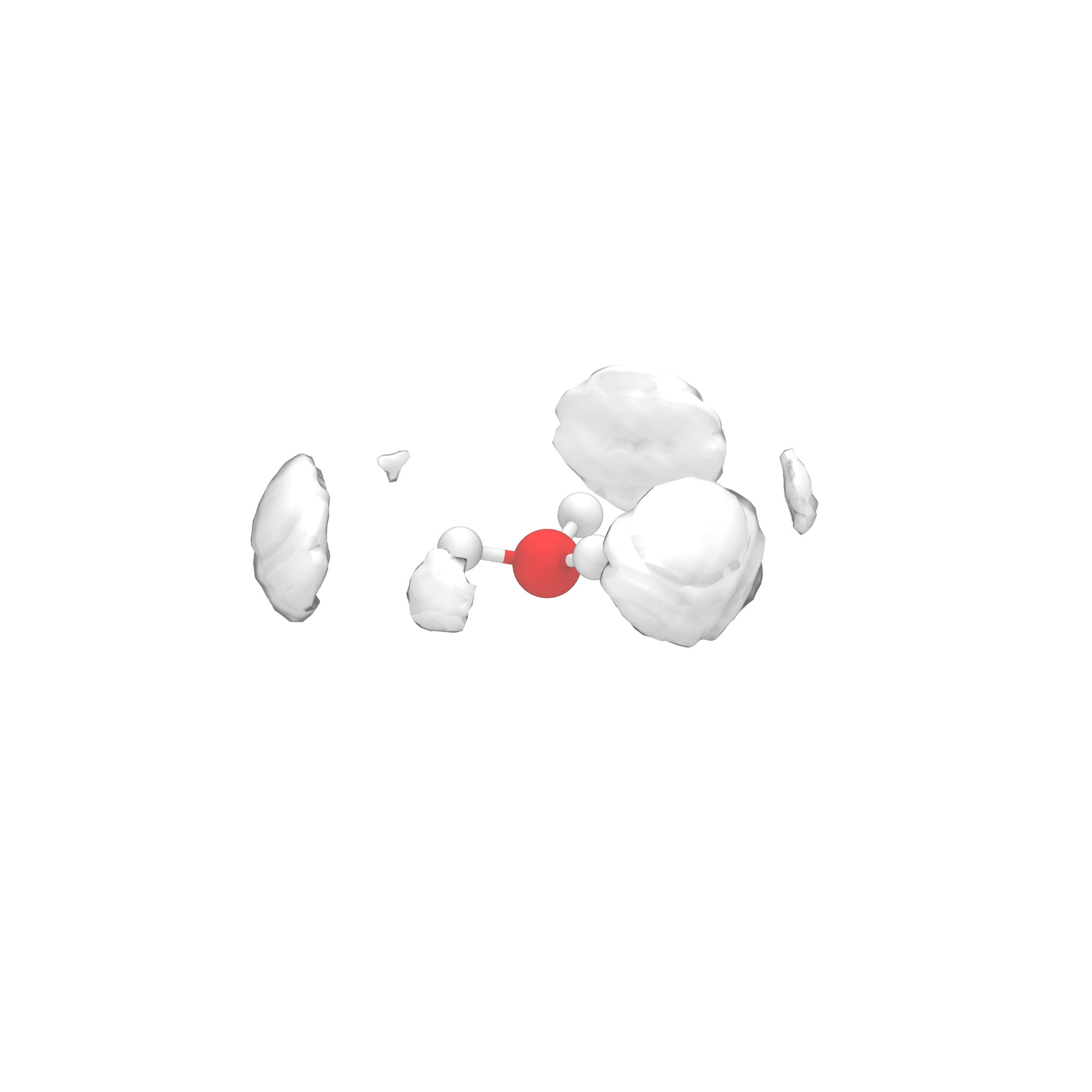}
    \end{minipage}\\
    \begin{minipage}[t]{0.25\columnwidth}
    \includegraphics[trim=400 700 400 600,clip,width=1.0\textwidth]{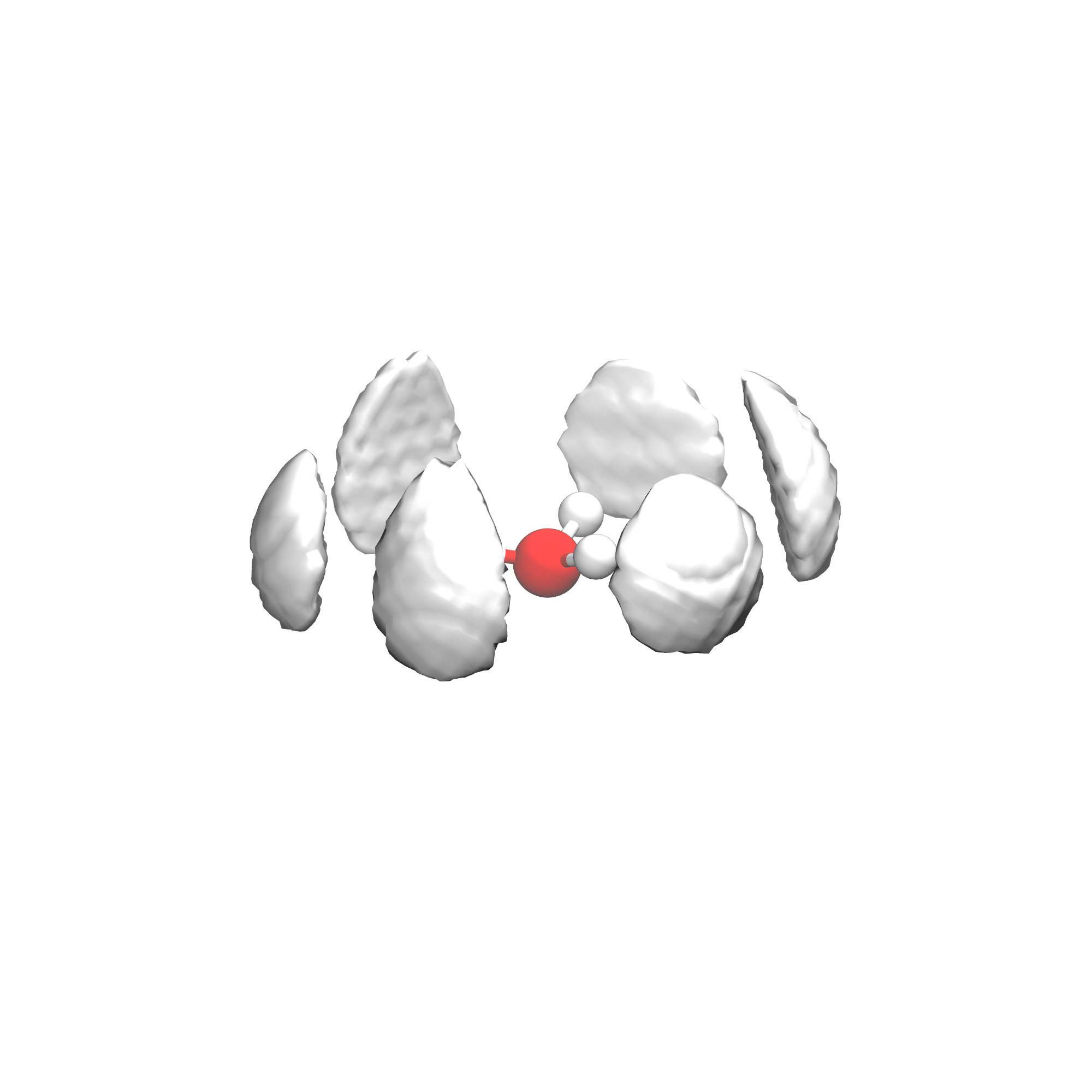}
    \end{minipage}
    \begin{minipage}[t]{0.25\columnwidth}
    \includegraphics[trim=400 700 400 600,clip,width=1.0\textwidth]{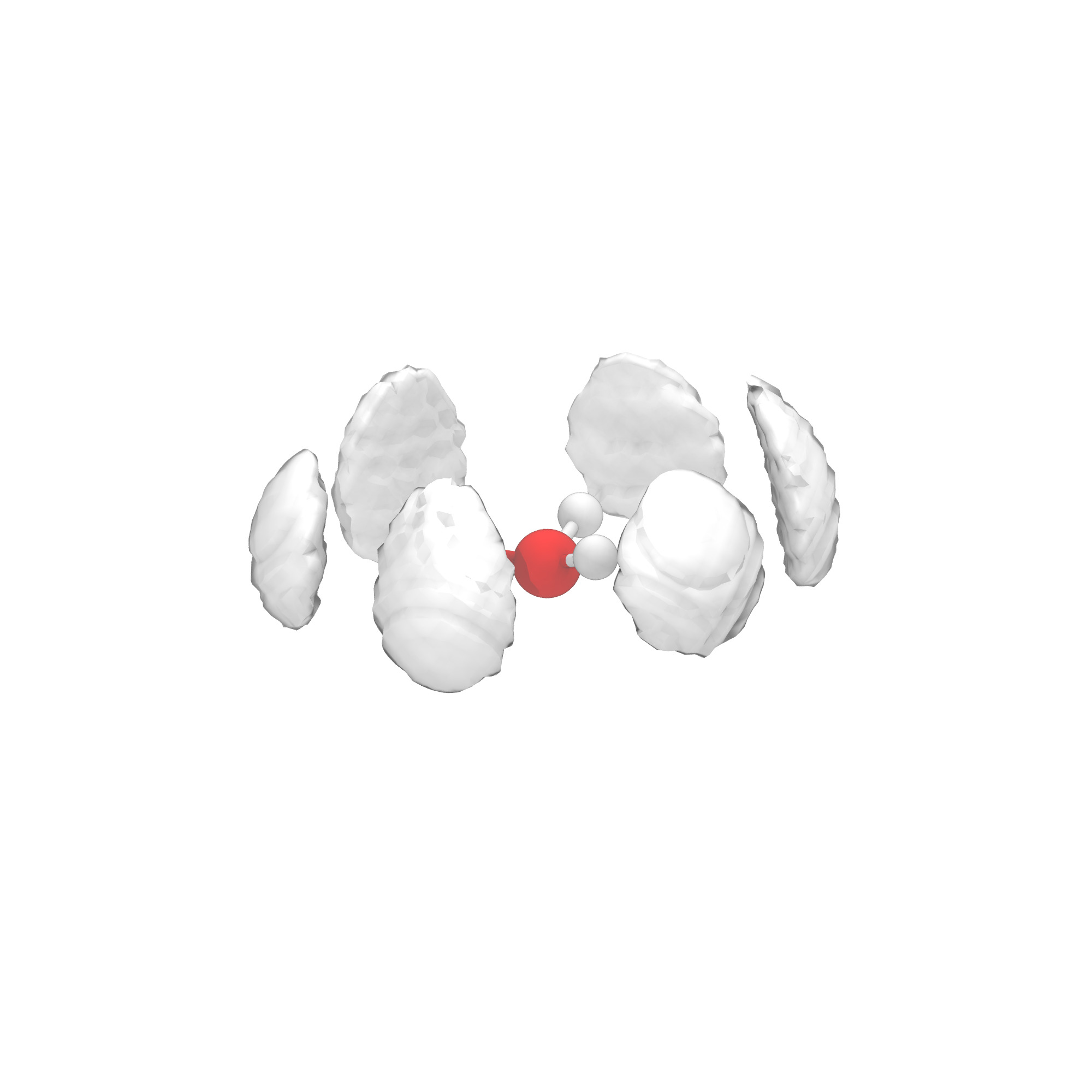}
    \end{minipage}\\
    \begin{minipage}[t]{0.25\columnwidth}
    \includegraphics[trim=400 500 400 450,clip,width=1.0\textwidth]{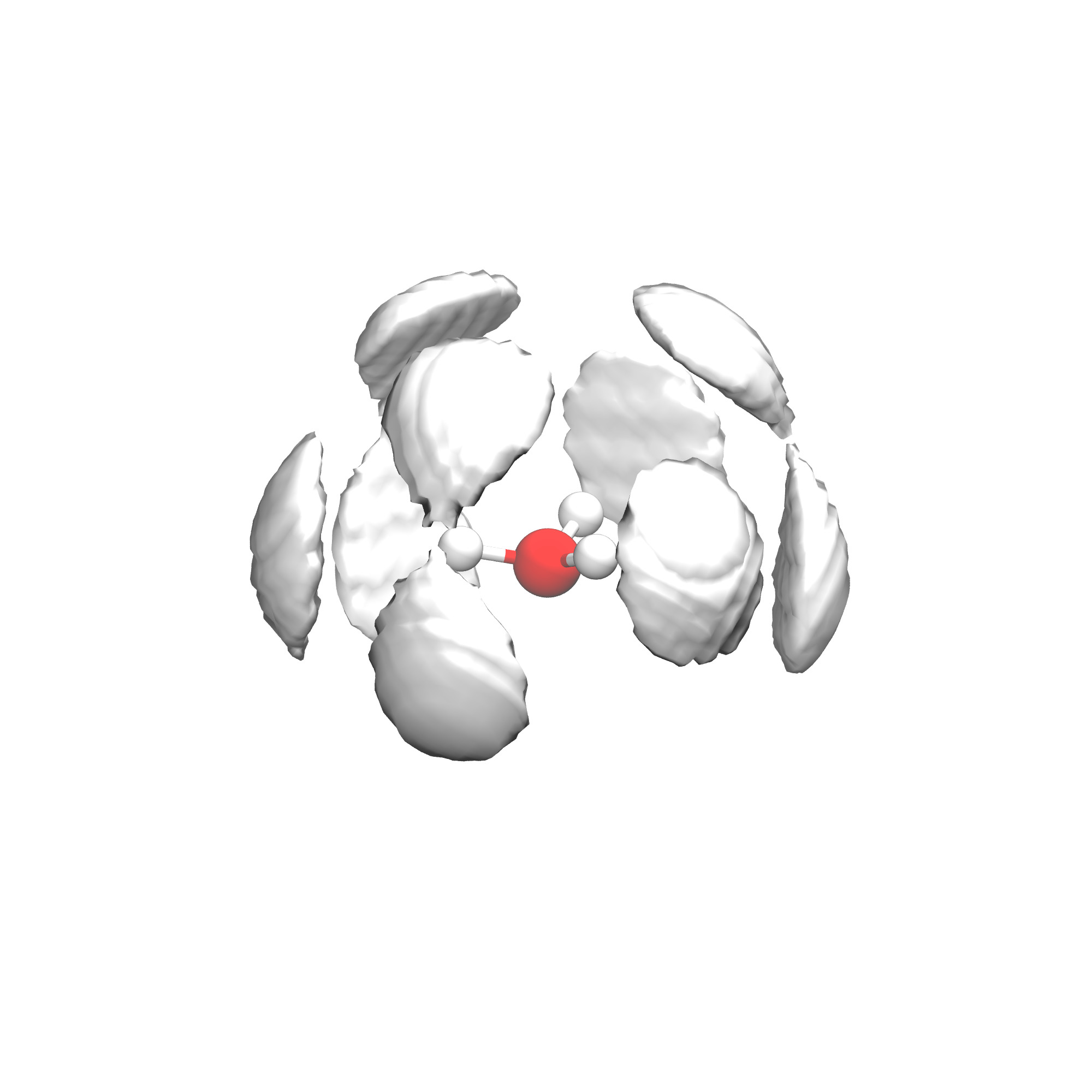}
    \end{minipage}
    \begin{minipage}[t]{0.25\columnwidth}
    \includegraphics[trim=400 500 400 450,clip,width=1.0\textwidth]{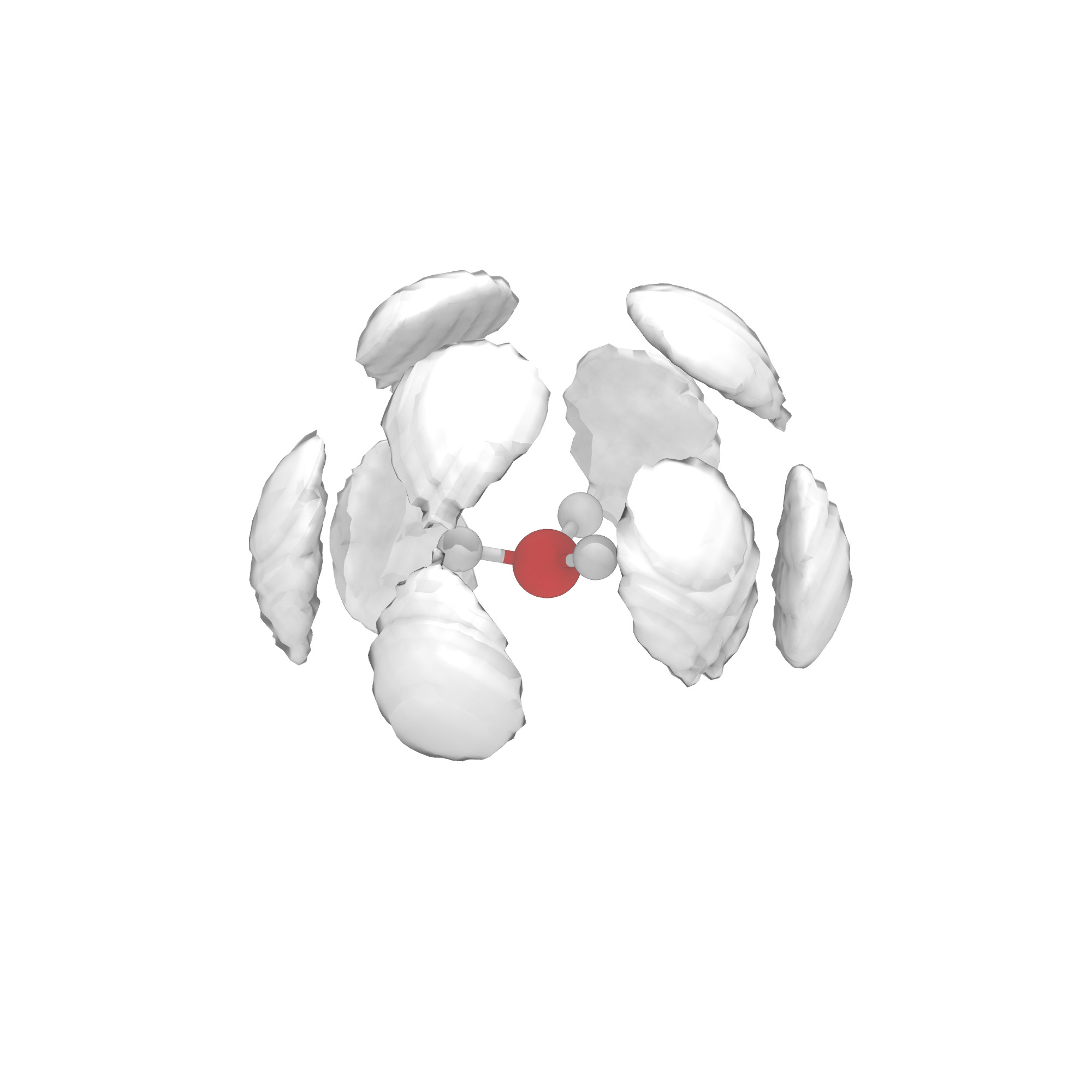}
    \end{minipage}\\
    \begin{minipage}[t]{0.25\columnwidth}
    \includegraphics[trim=400 440 400 440,clip,width=1.0\textwidth]{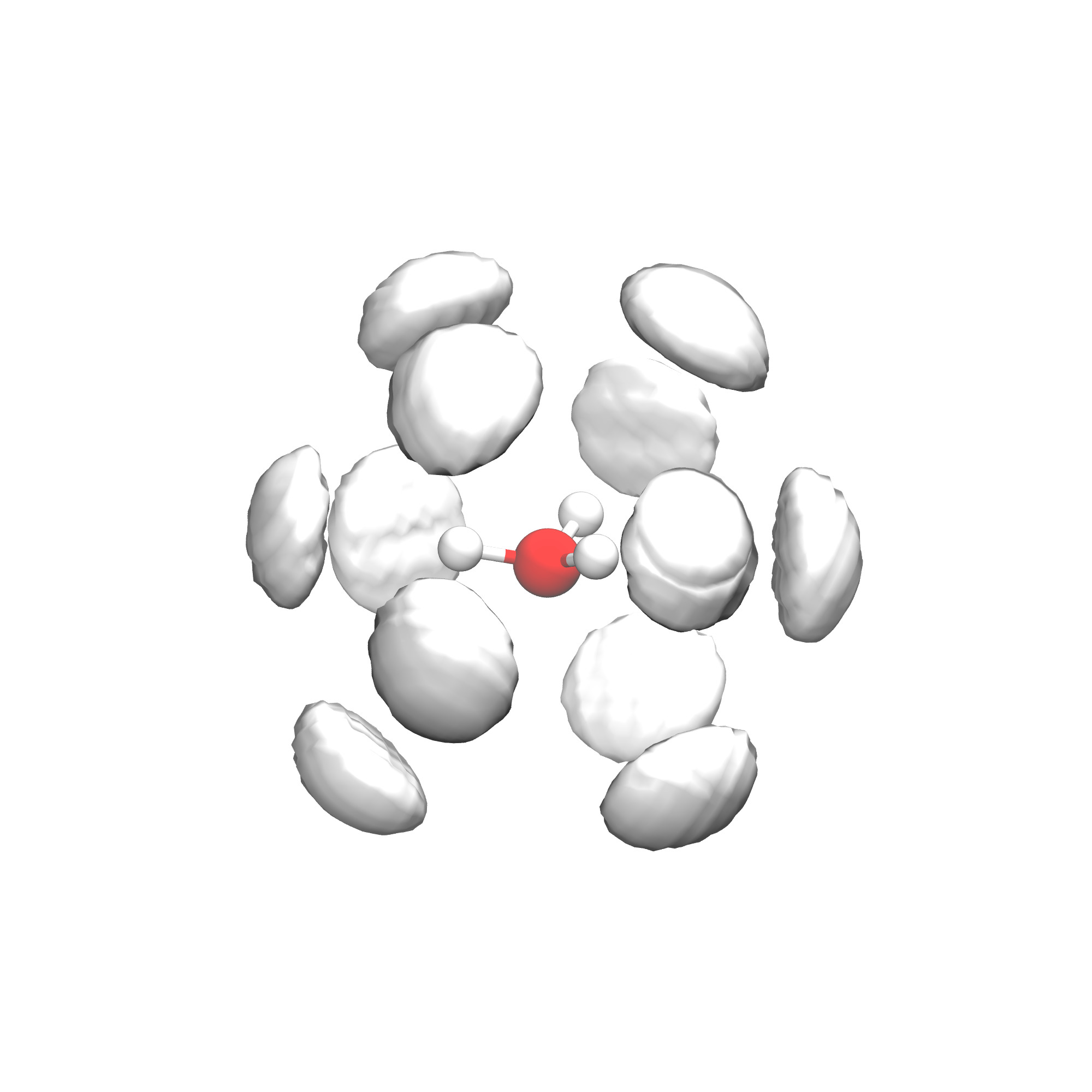}
    \end{minipage}
    \begin{minipage}[t]{0.25\columnwidth}
    \includegraphics[trim=400 440 400 440,clip,width=1.0\textwidth]{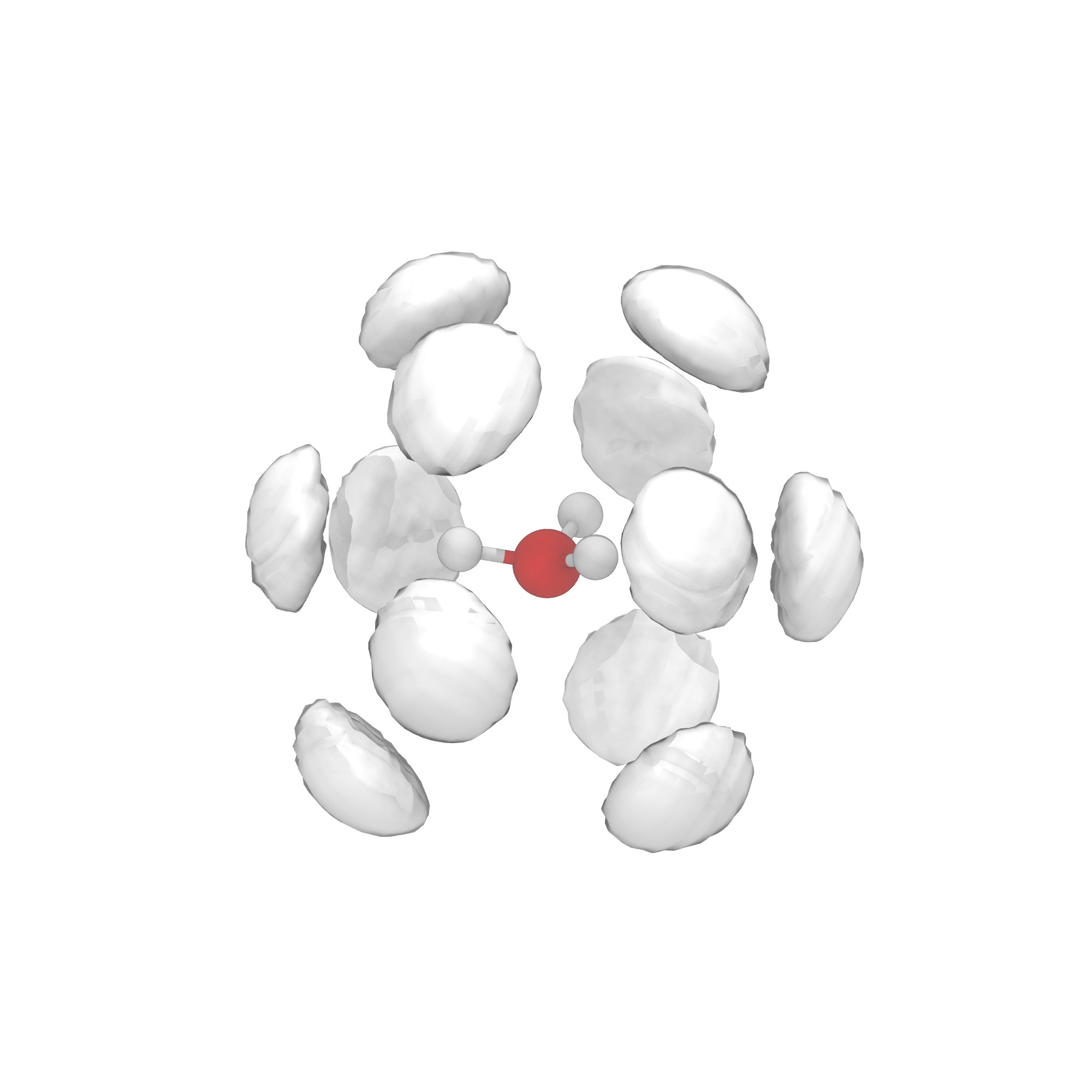}
    \end{minipage}
    \caption{Comparison of helium SDFs,
    see text,
    obtained from path integral 
    simulations with (from top to bottom) 1, 2, 4, 6, 10 and 14 helium atoms
    in the field of a frozen \hydronium{} configuration close to its minimum
    energy structure ('Minimum').
    Left: Energies obtained from the coupled cluster grid.
    Right: NNP evaluated at the coupled cluster grid points.
    The isovalue is set to 0.2$\cdot 10^{-3}$~1/bohr$^3$ in all shown cases.}
   \label{fig:Densitiesh3o+_min}
\end{figure}

In order to 
meaningfully
validate the quality of the converged NNPs, we 
recourse to the approach established 
in Ref.~\citenum{Uhl2017/10.1039/C7CP00652G}.
For a selected set of relevant fixed structures of the solute the evaluation of
the interaction potential on a grid is still feasible
even using the demanding CCSD(T)/AVTZcp reference method. 
Afterwards, helium can be simulated surrounding
the clamped solute structures by evaluating the 
solute-helium interaction on the grid using
the nearest neighbor approach in addition to the usual He-He
pair interactions.~\cite{Aziz1979/10.1063/1.438007}
This is significantly faster than the evaluation
of the NNP in continuum space by three orders of magnitude.
However, evaluating the NNPs takes 
on
the order of \SI{0.1}{\milli\second}
on a single core, 
while the CCSD(T) reference 
calculation
needs \SI{60}{\second} for
\hydronium{} and \SI{22}{\minute} for \zundelp{};
all reported timings were obtained on an Intel(R) Xeon(R) CPU E5-2630 v4 @ 2.20GHz.
Density functional theory calculations as described in section~\ref{sec:refval}, 
which might be used as an alternative method to determine the
interaction energy on-the-fly,
require
\SI{6}{\second} for \hydronium{} and \SI{50}{\second} for \zundelp{} on a single core.
To estimate
the influence of this 
grid-based
approach on the helium solvation structure,
simulations using the NNP in continuum space can be conducted.
For this procedure we chose two structures from the vacuum reference ensemble
of the \hydronium{} molecule. 
The first one is close to the minimum energy
configuration with slight deviations in the O-H bond length and
will be abbreviated as 'Minimum', while the second one
is a planar structure close to the transition state of the pseudo rotation and is 
called 
'Flat'.
For the \zundelp{} molecule three structures were selected from the reference
ensemble. 
The first is again closely related to the minimum
energy structure ('Minimum'), the second one is a planar structure ('Flat')
and in the third one the proton is in a very asymmetric position ('Asymm').
The interaction potential of these structures was evaluated 
as described in Sec.~\ref{sec:PISimulations} and afterwards applied to study
microsolvation of the solute by different numbers of helium atoms in the first step.
As a key property to compare the reference and NNP grid, we chose
to calculate the spatial distribution function (SDF) of helium, since
it is highly sensitive to small changes in the interaction energies.
The integral over space was normalized to the number
of helium atoms in order to compare the SDFs with different numbers
of helium atoms at the same isosurface value.

The comparison of the reference and the NNP grid
of hydronium microsolvated by helium is depicted in
Fig.~\ref{fig:Densitiesh3o+_min} for the Minimum configuration.
Overall, there are no noteworthy deviations between the SDFs
of the reference and the neural network
for all shown numbers of helium atoms.
Therefore, in the following we present the structural changes upon
increasing helium microsolvation of both grids combined.

\begin{figure}[t]
    \begin{minipage}[t]{0.25\columnwidth}
    \includegraphics[trim=400 400 400 400,clip,width=1.0\textwidth]{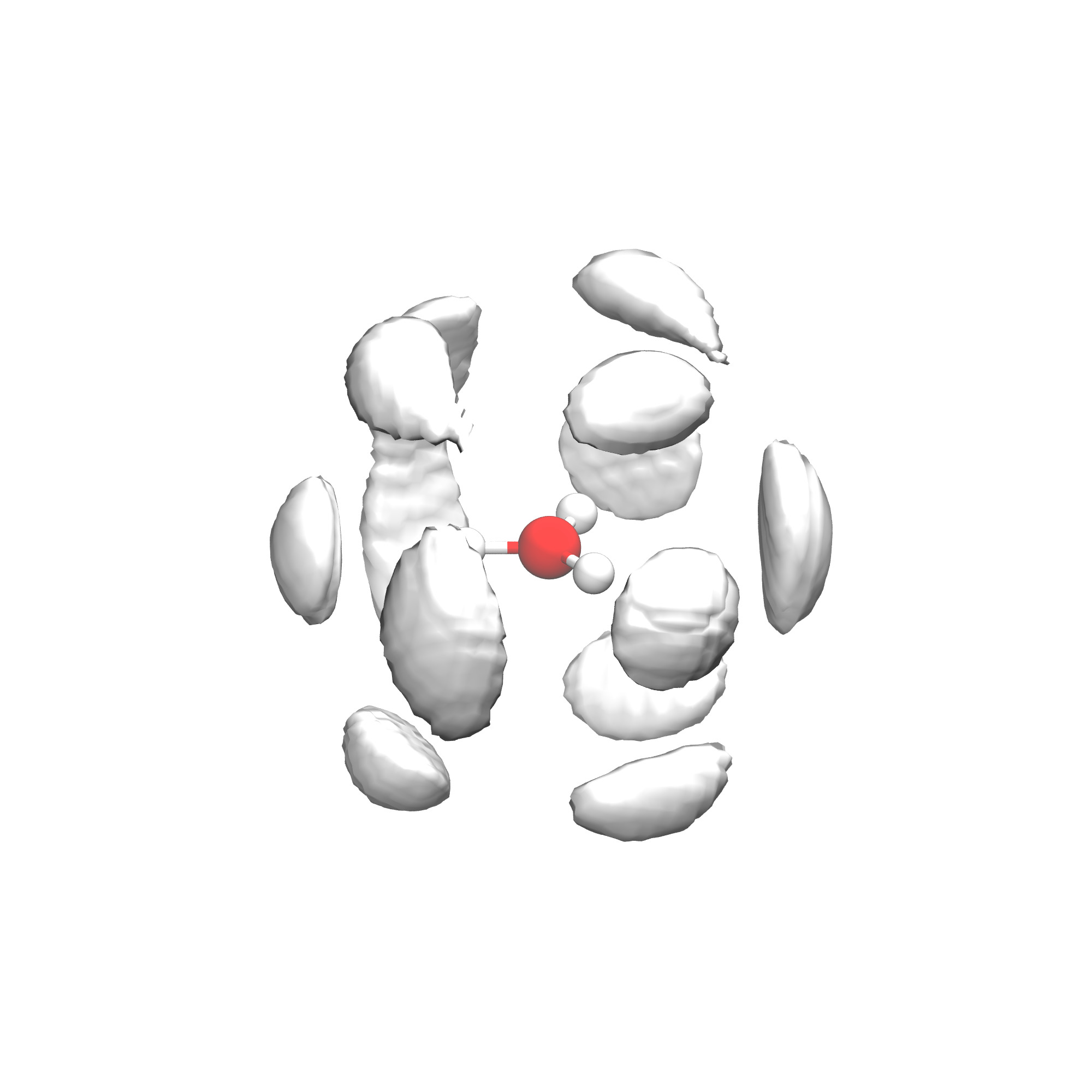}
    \end{minipage}
    \begin{minipage}[t]{0.25\columnwidth}
    \includegraphics[trim=400 400 400 400,clip,width=1.0\textwidth]{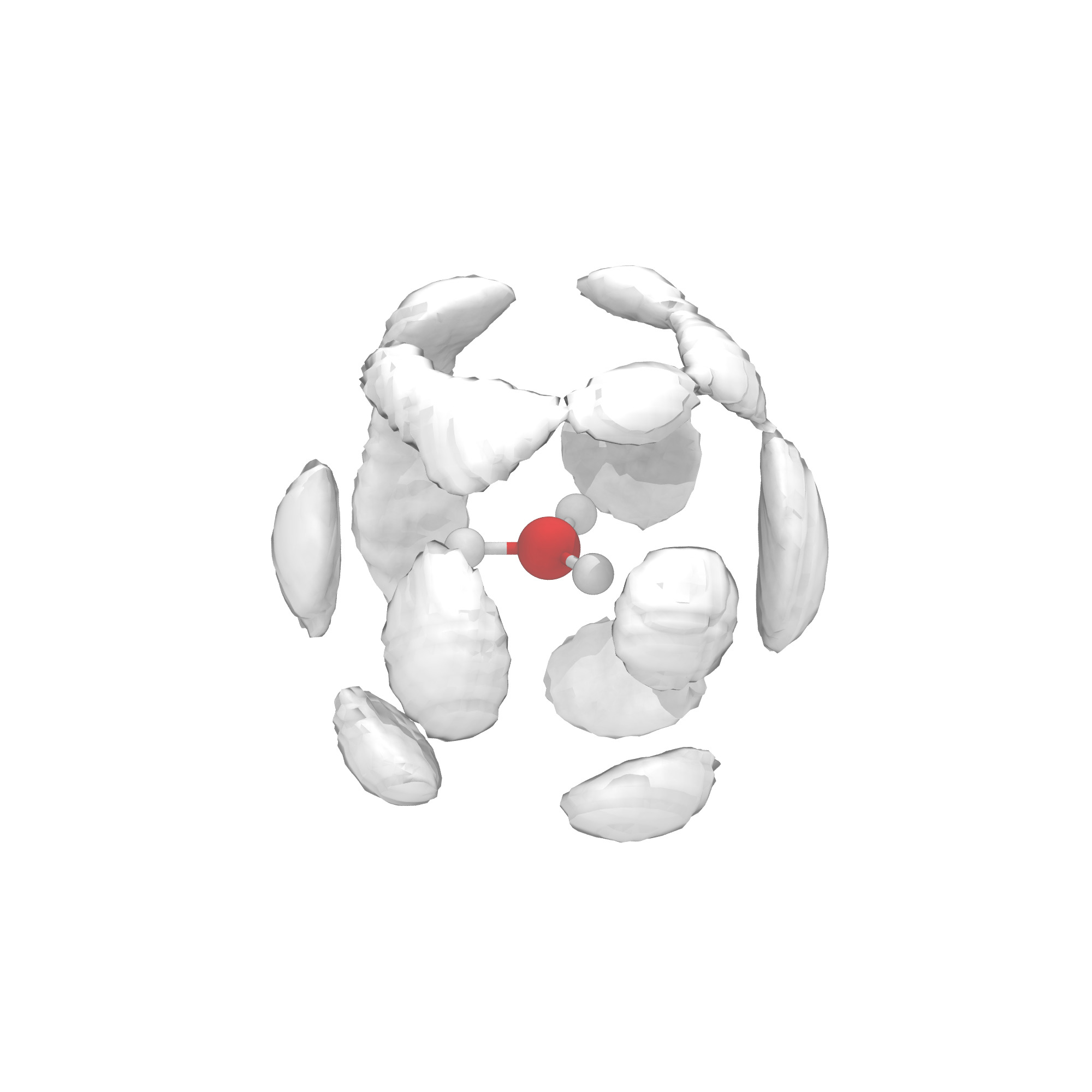}
    \end{minipage}\\
    \begin{minipage}[t]{0.25\columnwidth}
    \includegraphics[trim=400 400 400 400,clip,width=1.0\textwidth]{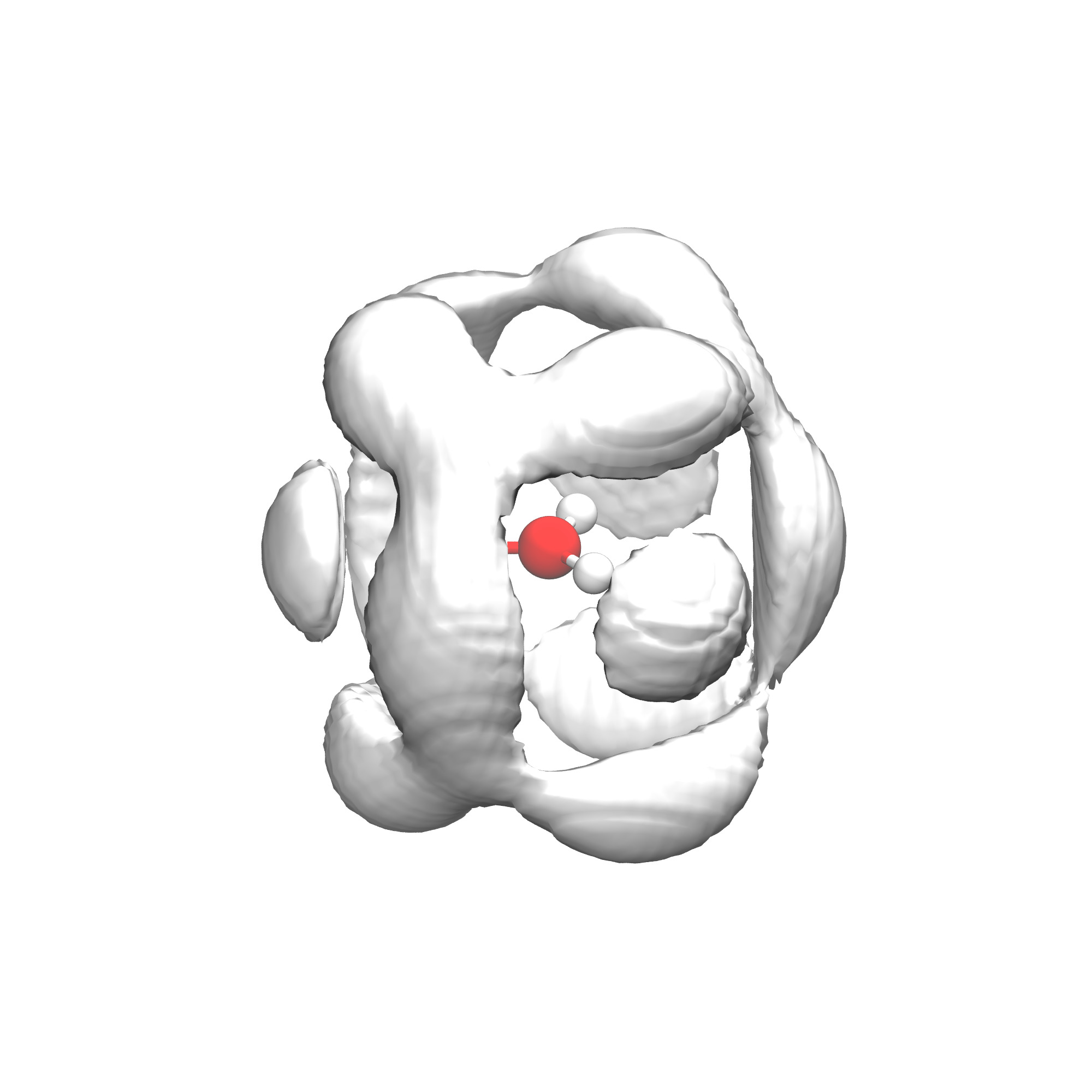}
    \end{minipage}
    \begin{minipage}[t]{0.25\columnwidth}
    \includegraphics[trim=400 400 400 400,clip,width=1.0\textwidth]{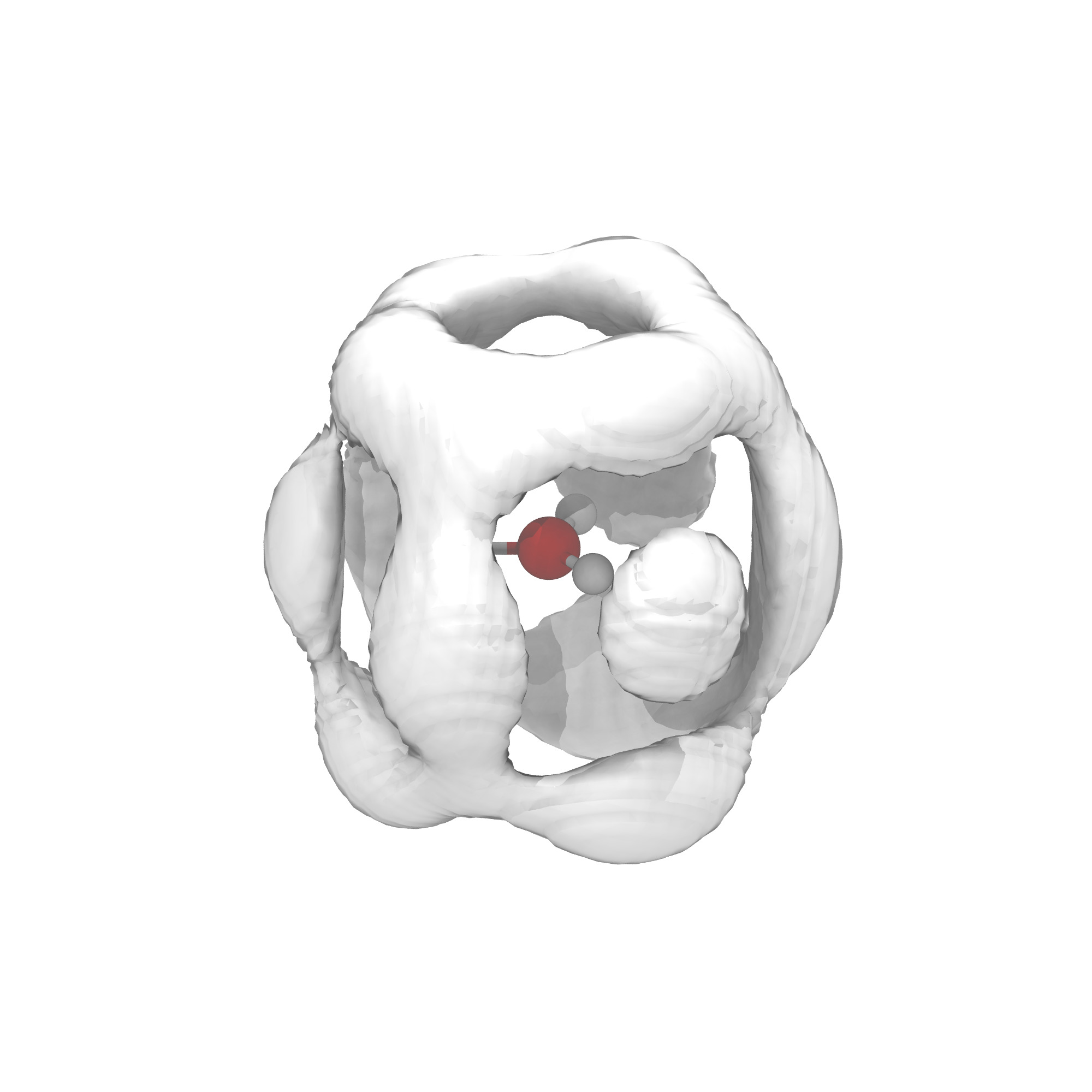}
    \end{minipage}
    \caption{Comparison of helium SDFs obtained from path integral 
        simulations with 14 helium atoms
        in the field of a frozen \hydronium{} configuration in a flat orientation ('Flat').
        Left: Energies obtained from the coupled cluster grid.
        Right: NNP evaluated at the coupled cluster grid points.
        Top: isovalue = 0.2$\cdot 10^{-3}$~1/bohr$^3$.
        Bottom: isovalue = 0.08$\cdot 10^{-3}$~1/bohr$^3$; note that the SDFs have 
        been normalized to the number of helium atoms.}
    \label{fig:Densitiesh3o+_flat}
\end{figure}

Microsolvation with one helium atom results in trapping the helium atom
inside one of the interaction wells that are located about 2~\AA{} away
from the hydrogen atoms in the direction of the O-H bonds.
During the simulations we did not observe hops between the three
almost equally deep wells and the appearance of three 
volumes of accumulated density (VAD) is only a result of
sampling with multiple MC walkers with different
random number seeds
in agreement with previous findings.~\cite{Uhl2017/10.1039/C7CP00652G}
The SDFs of two and three (not shown) helium atoms are therefore
almost identical to those with one atom.
When a fourth helium is added, however, it populates one of the three spatial
regions in between two interaction wells in order
to maximize its helium-helium interactions.
Again, the VADs appear in all three possible positions only due
to averaging over independent MC walkers.
For five (not shown) and six helium atoms these positions are saturated stepwise.
Addition of further helium atoms extends the solvation
to the region on top of the hydronium cation (in our chosen reference frame as
depicted in Fig.~\ref{fig:Densitiesh3o+_min}) as shown for ten helium atoms.
This region still features attractive interaction energies and
is in closer proximity to the hydrogen atoms than to the oxygen atom.
The helium atoms in between the interaction wells are pushed 
slightly out of the direct connection of the wells
and as a result, the VADs in the wells are elongated in the perpendicular
direction to the oxygen-hydrogen bonds in order to
increase the helium-helium interaction.
Microsolvation with 14 helium atoms results in a closed
first hydration shell also in the region closer to the
oxygen atom that features smaller interaction energies.
As already stated, the two grids yield almost identical
SDFs for all studied numbers of helium atoms and it can
therefore be concluded that the NNP describes the 
microsolvation of this hydronium configuration quantitatively
compared to the CCSD(T)/AVTZcp reference.

For the sake of clarity, we do not present this comparison for
all selected solute structures, since their general dependence on
the number of helium atoms is similar and the results obtained
with the NNP grids are very close to the reference in most cases.
Instead, we present all helium SDFs in 
Section~II of
the SI and concentrate here
on situations where we do not
observe perfect agreement between the reference and the NNP.
The first of these cases is the microsolvation of
the flat hydronium structure by 14 helium atoms as depicted
in Fig.~\ref{fig:Densitiesh3o+_flat}.
Although up to this number of helium atoms we observe
again very good agreement between the two grids as can be
seen in 
Sec.~II.A in
the SI, the shown SDFs feature small differences.
The region in between the upper VADs for the larger
isovalue depicted in the figure is connected for the NNP grid, while
it is not for the reference grid.
However, this is merely a consequence of the high sensitivity to the isovalue 
chosen to represent the SDFs. 
For smaller isovalues
this connection also appears for the reference grid and
the differences between the grids are again very small
as shown in the lower panel of Fig.~\ref{fig:Densitiesh3o+_flat}.
This thus suggests that the NNP energy is slightly too
shallow in this particular region of the interaction potential.
Note that these findings are converged with respect to the simulation
length as presented in Sec.~II.C of the SI where we compare
the SDFs to simulations with ten times improved statistics.
In addition, flat hydronium structures like the one used for the grid
are very rare and make up only 3~\% of the 
vacuum reference ensemble.
We are therefore confident that this small difference in
relative energies is of minor 
importance
for the overall microsolvation of \hydronium{} by helium.

\begin{figure}[t]
    \begin{minipage}[t]{0.25\columnwidth}
    \includegraphics[trim=450 600 450 600,clip,width=1.0\textwidth]{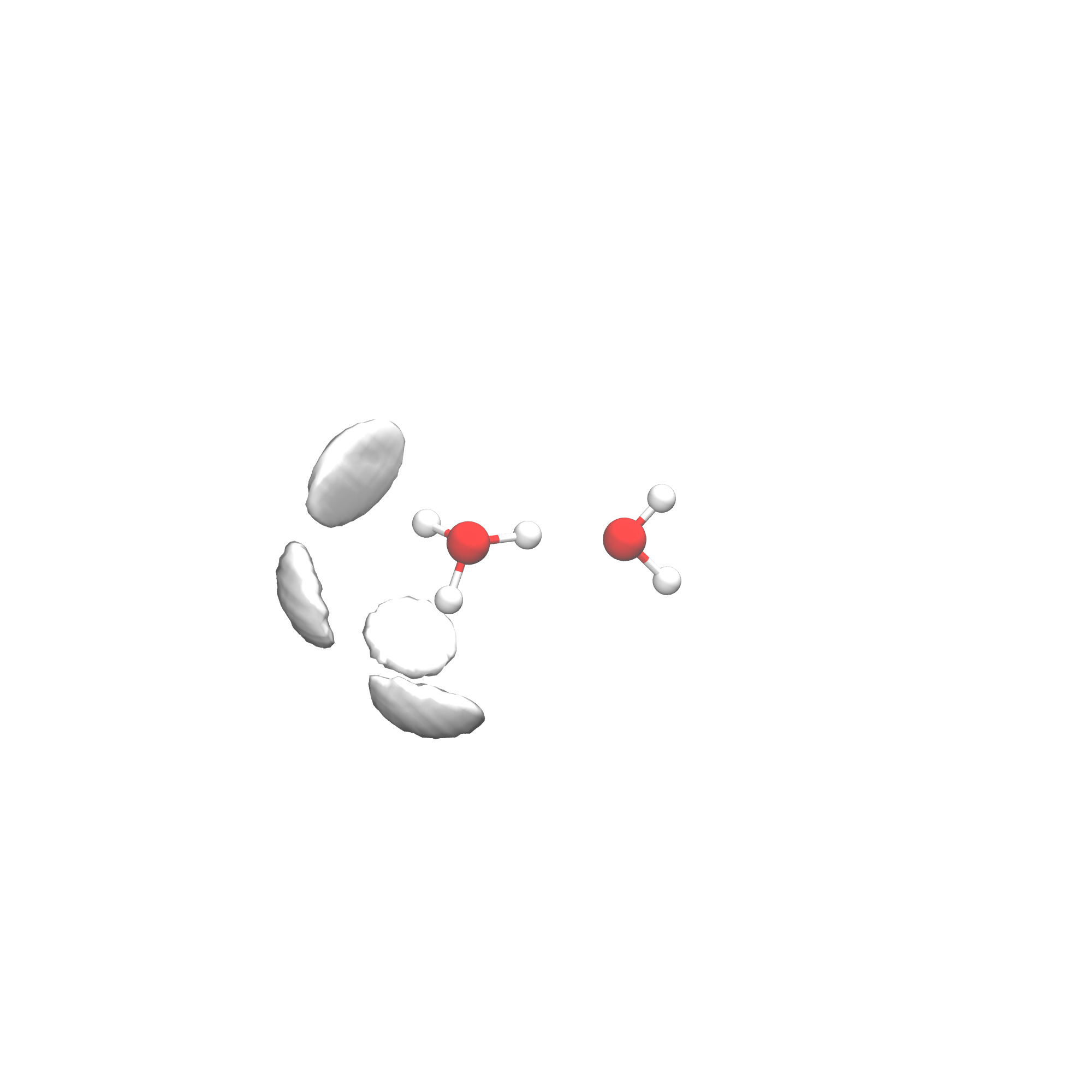}
    \end{minipage}
    \begin{minipage}[t]{0.25\columnwidth}
    \includegraphics[trim=450 600 450 600,clip,width=1.0\textwidth]{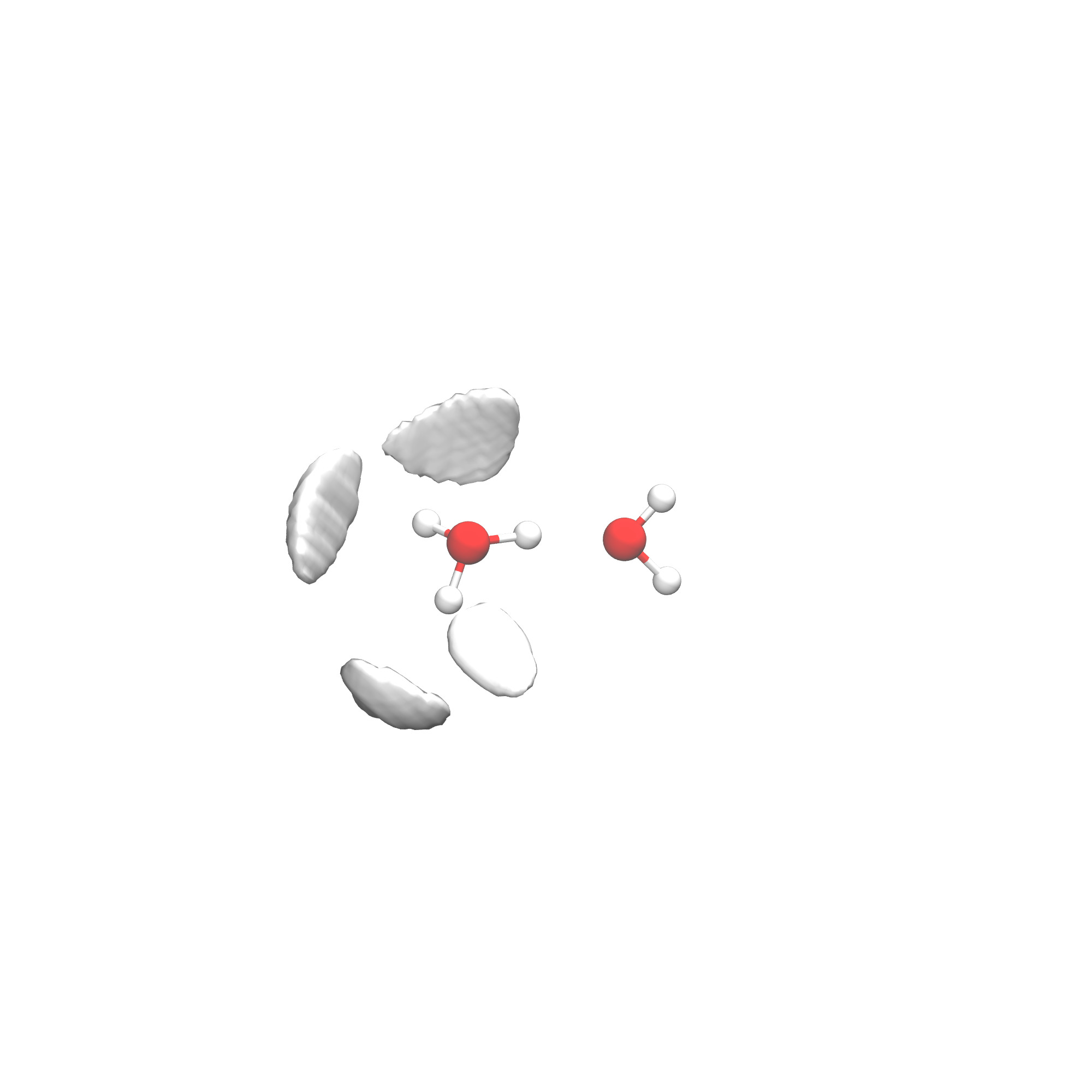}
    \end{minipage}\\
    \begin{minipage}[t]{0.25\columnwidth}
    \includegraphics[trim=450 600 450 600,clip,width=1.0\textwidth]{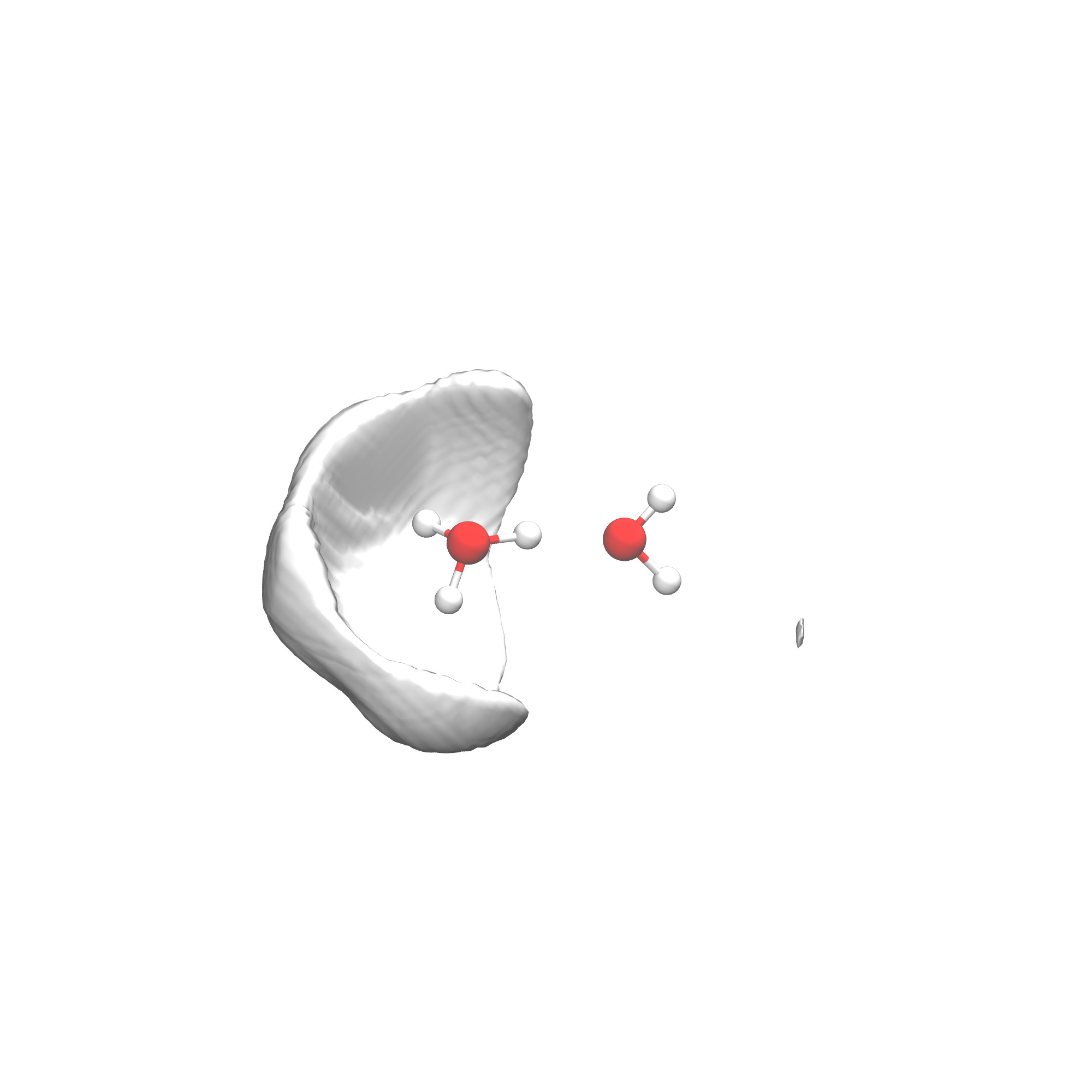}
    \end{minipage}
    \begin{minipage}[t]{0.25\columnwidth}
    \includegraphics[trim=450 600 450 600,clip,width=1.0\textwidth]{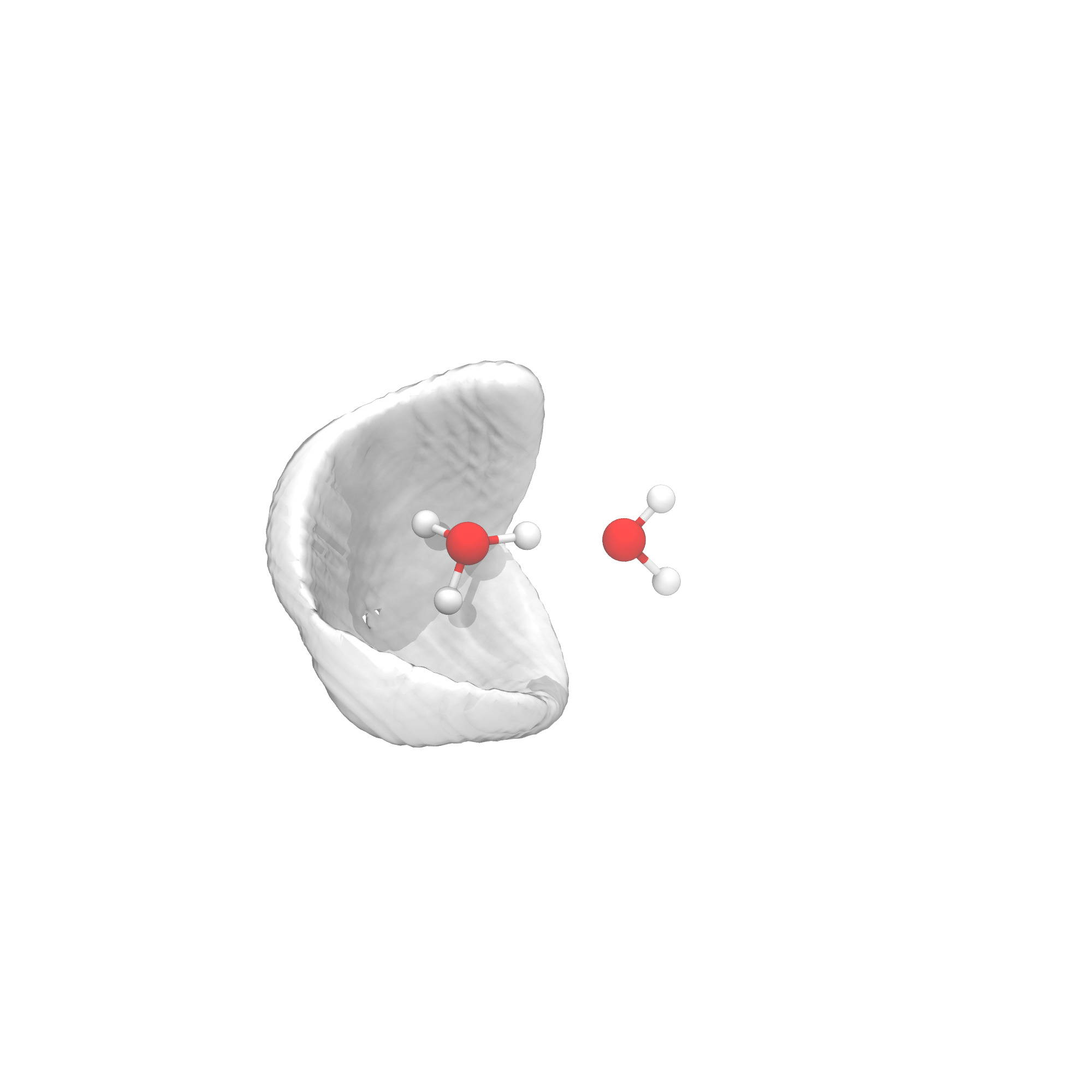}
    \end{minipage}
    
    \vspace{3em}
    \begin{minipage}[t]{0.25\columnwidth}
    \includegraphics[trim=450 600 450 600,clip,width=1.0\textwidth]{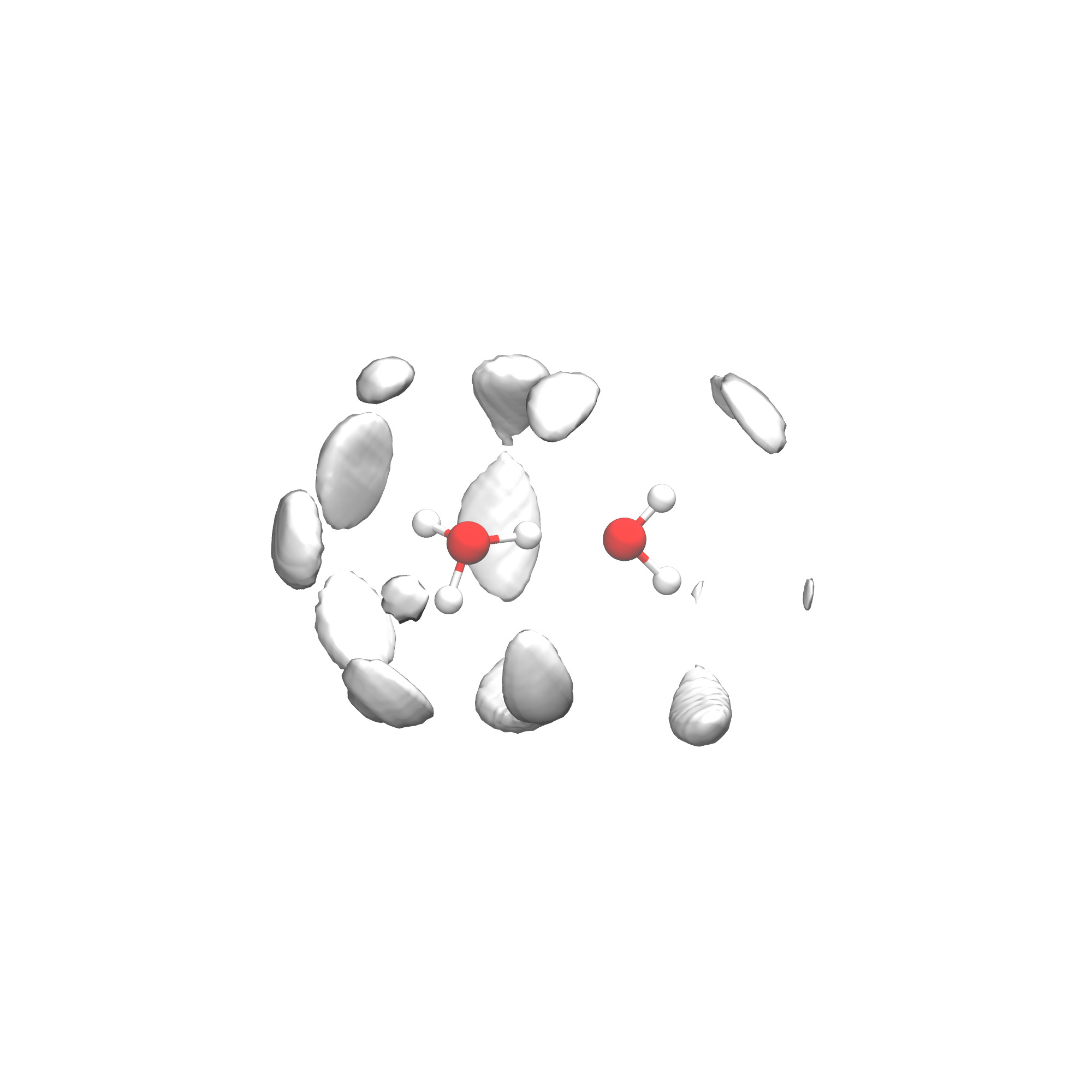}
    \end{minipage}
    \begin{minipage}[t]{0.25\columnwidth}
    \includegraphics[trim=450 600 450 600,clip,width=1.0\textwidth]{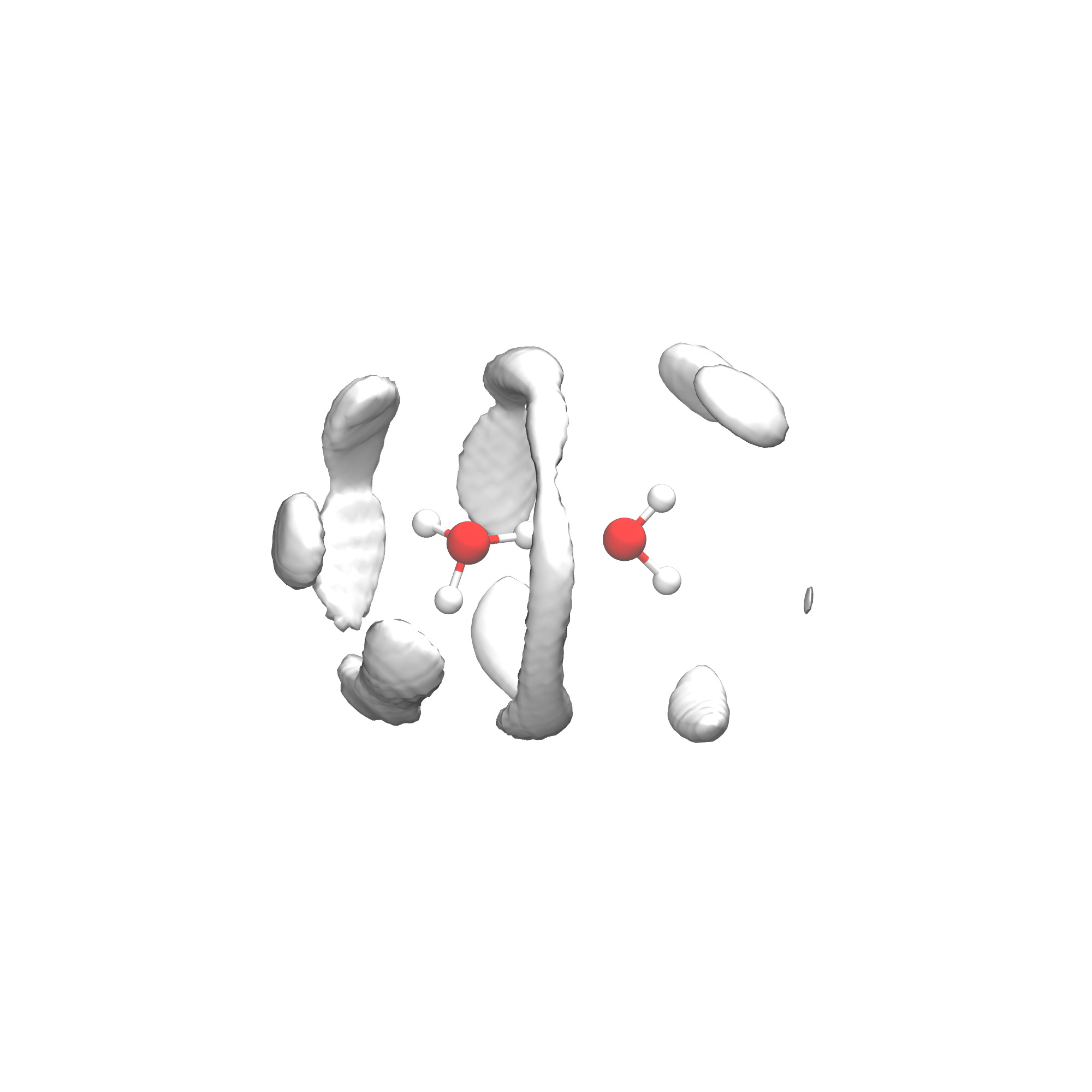}
    \end{minipage}\\
    \begin{minipage}[t]{0.25\columnwidth}
    \includegraphics[trim=450 550 450 550,clip,width=1.0\textwidth]{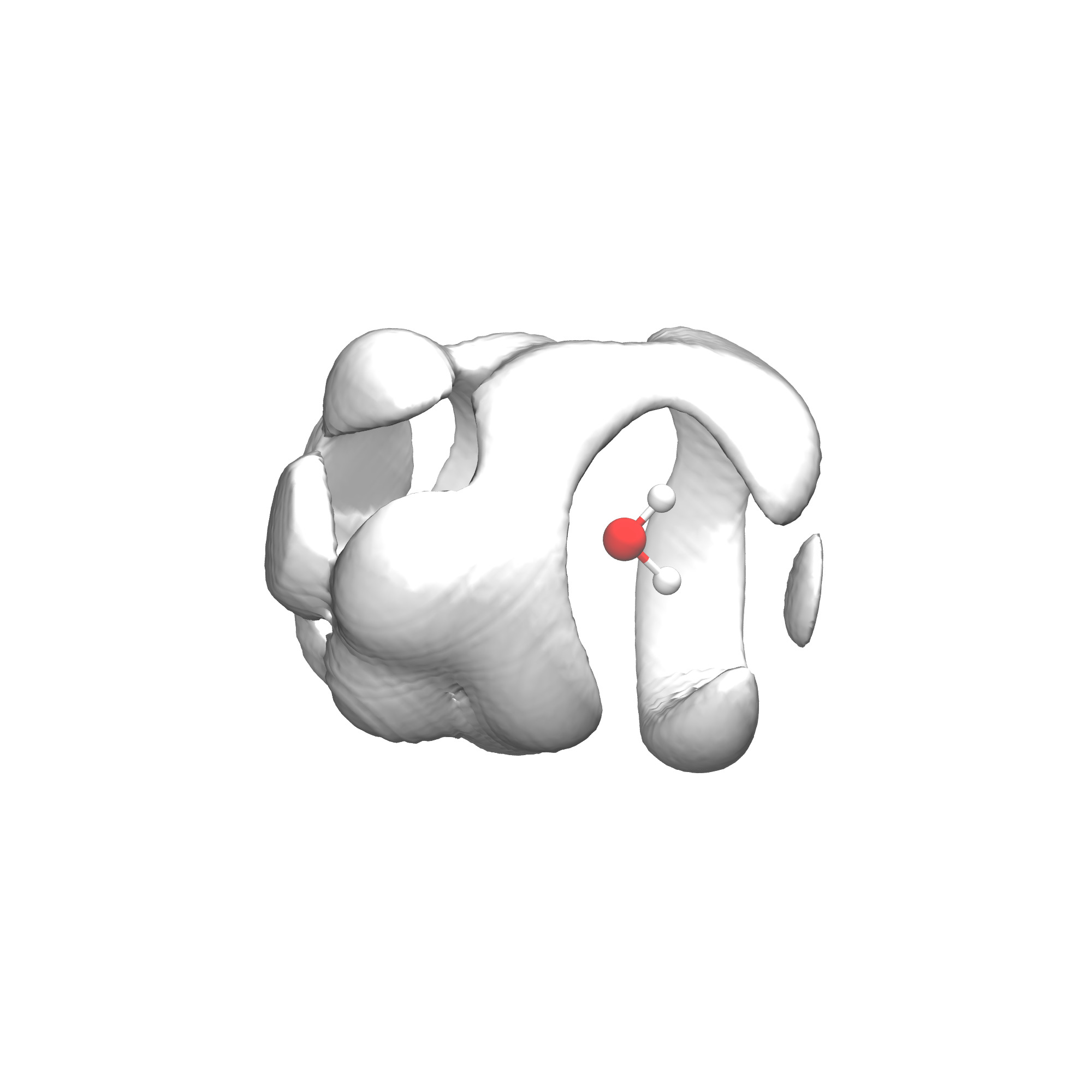}
    \end{minipage}
    \begin{minipage}[t]{0.25\columnwidth}
    \includegraphics[trim=450 550 450 550,clip,width=1.0\textwidth]{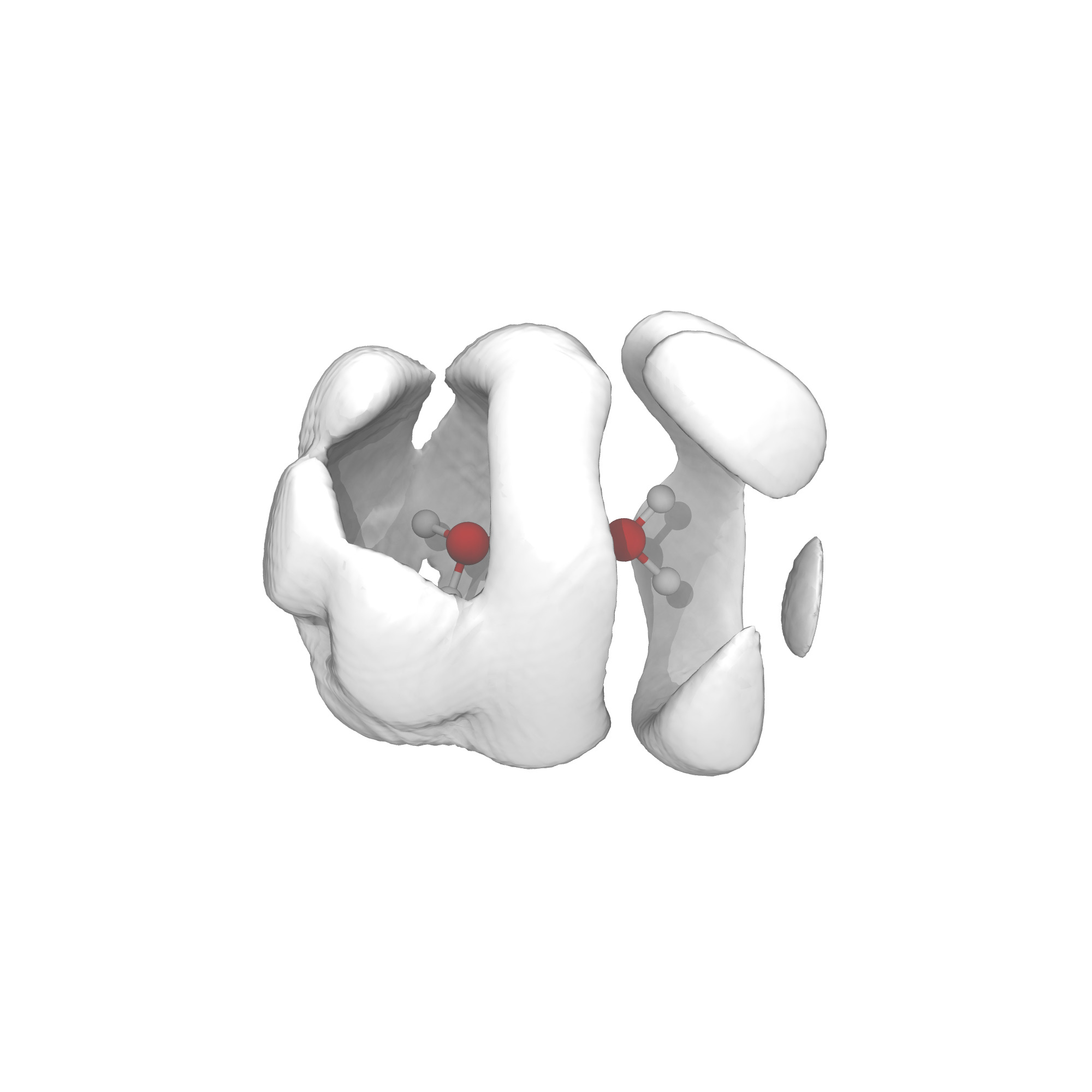}
    \end{minipage}
    \caption{Comparison of helium SDFs obtained from path integral 
        simulations with 4 (two top panels) and 14 (two bottom panels) 
        helium atoms in the field of a frozen \zundelp{}
        configuration in an asymmetric proton transfer situation ('Asymm').
        Note that configuration with a similar value of the proton transfer
        coordinate only make up for around 1-2~\% of all configurations.
        Left: Energies obtained from the coupled cluster grid.
        Right: NNP evaluated at the coupled cluster grid points.
        The isovalue for both numbers of helium atoms in the
        upper representation was set to 0.2$\cdot 10^{-3}$~1/bohr$^3$, while for the
        lower representation it was set to 0.035$\cdot 10^{-3}$~1/bohr$^3$;
        note that the SDFs have 
        been normalized to the number of helium atoms.
        \label{fig:Densitiesh5o2+_flat}}
\end{figure}

Microsolvation of the Zundel cation by helium is equally well described
by the NNP for almost all numbers of helium atoms
that have been 
studied for the three selected \zundelp{} structures.
The reader is again referred to the SI,
see Section~II.A therein,
for a comprehensive comparison.
Again, we only present here the two cases where the SDFs obtained
on the NNP grid 
apparently
deviate from the reference as depicted in
Fig~\ref{fig:Densitiesh5o2+_flat}.
Actually, the selected \zundelp{} configuration with asymmetric proton sharing
is the only candidate where the SDFs of the NNP grid and the reference grid
partially disagree, specifically for four and 14 helium atoms.
While the reference grid for $n=4$ helium atoms provides two VADs in between the VADs of the
interaction wells as visible at the larger isovalue in Fig~\ref{fig:Densitiesh5o2+_flat}
on the side of the hydronium-like motif of the complex,
the minima on the NNP are too close and therefore
displace the remaining two helium atoms to the surrounding.
In addition, the part of the interaction potential around the
excess proton is shallower on the NNP grid, similarly as observed
for the flat \hydronium{} configuration. 
This results in connected VADs in the
case of microsolvation with 14 helium atoms, although the position of the remaining
VADs is well captured by the NNP.
The discrepancies are again eliminated when the SDFs
are visualized at smaller isovalues as shown in Fig~\ref{fig:Densitiesh5o2+_flat}
which gives us confidence that the microsolvation of this species
is essentially described correctly by the NNP.
In addition, the excess proton in this configuration 
features a comparable O--H bond length as for the dangling
hydrogen and is, thus, in an unfavorable configuration.
Actually, structures with a similar proton sharing situation only account
for 1-2\% of the vacuum ensemble and are therefore again of minor importance.
We therefore conclude that also the microsolvation of the Zundel cation by
helium is captured correctly by the converged NNP.

\subsection{Bulk Helium Solvation}
\label{sec:bulk}

\begin{figure}[t]
    \includegraphics[width=1.0\linewidth]{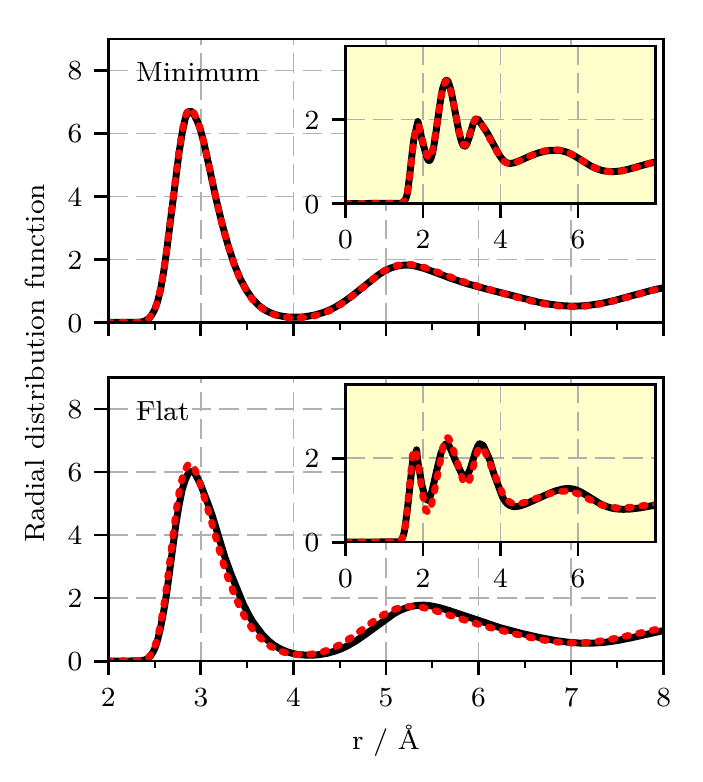}
    \caption{Radial distribution function for oxygen-helium (main) and
    hydrogen-helium (inset) for frozen \hydronium{} configurations in bulk helium 
    in a selected orientation close to the minimum energy structure 
    ('Minimum', top) and in a flat orientation ('Flat', bottom) 
    centered in a truncated octahedron periodic supercell
    hosting in addition 98 He atoms.
    Black, solid: Energies obtained from the coupled cluster grid.
    Red, dotted: NNP evaluated at the coupled cluster grid points.
    \label{fig:w1h_rdf}}
\end{figure}

In order to probe the quality of the developed NNPs
for solvation of
these protonated water clusters
in bulk helium, we performed bulk simulations
using the same nearest neighbor approach as employed in the previous section
which is described in detail in Sec.~\ref{sec:PISimulations}.
In addition to the SDFs, we also chose to calculate the
RDFs between the solute atoms and helium
as defined in Ref.~\citenum{TuckermanBook},
since they provide 
easier insight into the solute-induced structuring of the solvent in terms of
the different solvation shells.
The oxygen- and hydrogen-helium RDFs of the \hydronium{} molecule in bulk helium are
shown in Fig.~\ref{fig:w1h_rdf} for both 
reference structures, Minimum and Flat.
Note that we present only one representative H-He RDF
for both reference structures, namely in both cases the RDF
of the left hydrogen atom defined according to the reference
frame in Figs.~\ref{fig:Densitiesh3o+_min} and~\ref{fig:Densitiesh3o+_flat}.
The other RDFs are very similar and feature the same agreement between
NNP and CCSD(T) grid.

The first helium solvation shell has its maximum at roughly 2.9~\AA{} apart from
the oxygen atom for both structures, while the second shell peaks at 5.3~\AA{}.
In between 
the two peaks
the O-He RDFs 
decay
almost to zero, suggesting that
the first solvation shell is very well structured and that helium atoms do
not migrate much between the shells.
This conclusion is also supported by the sharp and intense peak
of the first shell and is 
consistent with the
anisotropic helium distributions
around solutes from earlier predictions in pioneering studies on this 
subject
using very simple He-solute interaction models.~\cite{Barnett1993/10.1063/1.465455,
Kwon1996/10.1063/1.470929}
It is also consistent with the general notion of “frozen” helium atoms
that are tightly bound to 
certain
impurities.~\cite{Kwon2000/10.1063/1.1310608, 
Paolo2006/10.1140/epjd/e2006-00184-1,Coccia2007/10.1063/1.2712437,Galli2011/10.1021/jp200617a}
The fine structure of the first shell can be better seen in the H-He
RDFs shown in the inset of the figure for one of the hydrogen atoms.
They feature three pronounced peaks in line with the well structured VADs
observed for the microsolvated solute configurations.
As in the previous section, the NNP grids yield almost identical RDFs 
compared to the reference 
CCSD(T)
grid for both structures.
In case of the flat configuration that accounts only
for 3\% of the ensemble, there are small deviations, but the
overall bulk solvation pattern according to the reference grid is reproduced
very well by the NNP.
Note that the observed differences are in the same order than the basis set error
introduced by not using the complete basis set limit as shown 
in Section~I.A of the SI.

\begin{figure}[th!]
    \includegraphics[width=1.0\linewidth]{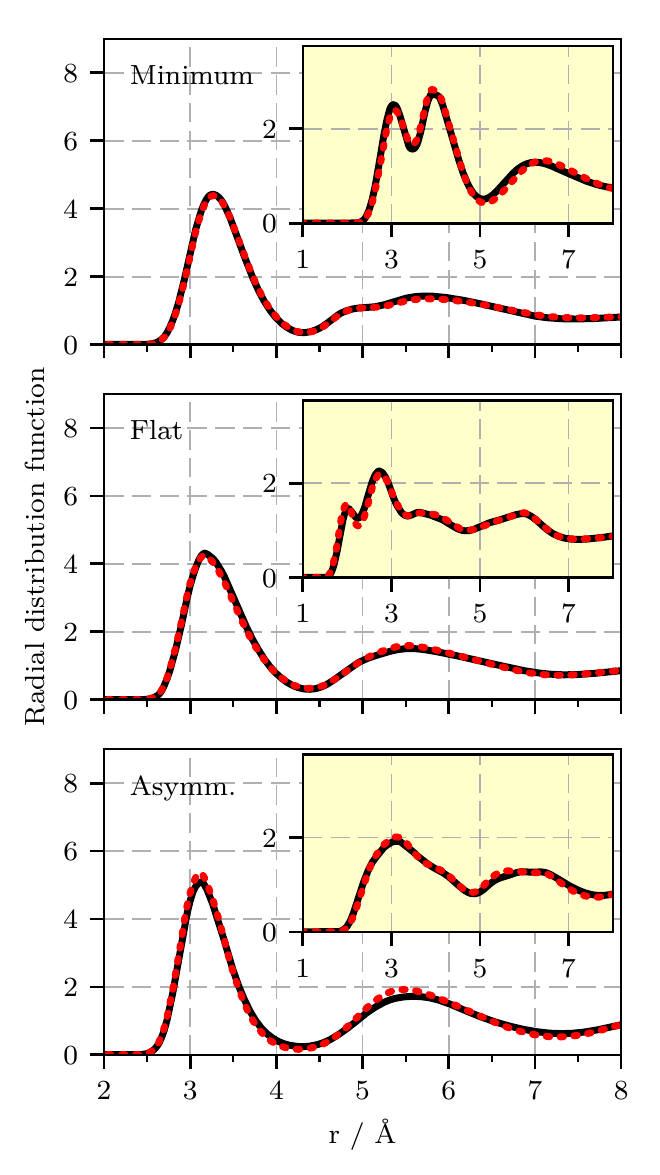}
    \caption{Radial distribution function for oxygen-helium (main) and
        hydrogen-helium (inset) for frozen \zundelp configurations in bulk helium 
        in a selected orientation close to the minimum energy structure 
        ('Minimum', top), in a flat orientation ('Flat', middle)
        as well as in an asymmetric proton transfer situation 
        ('Asymm', bottom) centered in a truncated octahedron periodic supercell
        hosting in addition 88 He atoms.
        Black, solid: Energies obtained from the coupled cluster grid.
        Red, dotted: NNP evaluated at the coupled cluster grid points.
        \label{fig:w2h_rdf}
        }
\end{figure}

As expected from the RDFs, the SDFs of the structure close to the minimum
energy configuration are in perfect agreement 
(only shown in Section~II.B of the SI) and
compare in their first solvation shell well with the microsolvation shell 
obtained with 14 helium atoms.
The SDFs of the flat configuration 
(shown in Section~II.B of the SI),
feature small deviations 
that can again be explained by similar reasons as before:
Some parts of the interaction potential are slightly shallower
on the NNP grid and thus yield pronounced VADs already at higher isovalues
compared to the reference grid.
The same arguments as presented in the previous section in combination
with the good agreement of the RDFs gives us confidence that the
solvation by bulk helium of even rare 
event 
structures like the flat configuration, being
close to the transition state of the pseudorotation of the \hydronium{}
cation, is correctly described by our He-hydronium NNP.

Finally, let us 
focus on the \zundelp{} cation and its solvation
by bulk helium. 
The comparison of the O-He and H-He RDFs for the three 
selected configurations is
depicted
in Fig.~\ref{fig:w2h_rdf}.
As before, we present only two representative RDFs
for each reference structure, namely the H-He RDF of the shared
proton for Minimum and the H-He RDFs of a selected dangling hydrogen
for Flat and Asymm. The O-He RDFs are shown for the
left oxygen atom according to the reference
frame in Fig.~\ref{fig:Densitiesh5o2+_flat} and
Sec.~II.A of the SI.
Again, the other RDFs are similar and yield the same level of agreement between
NNP and CCSD(T) grid.
Compared to the \hydronium{} molecule, the first O-He shell peaks at
larger distances of about 3.2-3.4~\AA{} and are also broader.
This implies a less tight binding in the first solvation shell compared
to the hydronium cation which is supported by the higher probability at the
first minimum in the RDFs compared to the \hydronium{} molecule.
Also our findings from Sec.~\ref{sec:res_nnfit} that the maximum of
the interaction potential between helium and the Zundel cation
is about \SI{2.5}{\kilo\joule\per\mole} lower is in accordance with
these less structured RDFs. 
As expected, the sharpest
first peak is observed for the structure with an asymmetric proton
which features the largest charge localization.
As already found for the smaller \hydronium{} molecule, the NNP grids provide
almost identical RDFs compared to the reference grids for all
three selected structures.
Also the SDFs of \zundelp{} in bulk helium support the conclusion that
the NNP is able to describe the helium-solute interaction with
almost perfect agreement to the reference as can be seen 
in Section~II.B of the SI.
Even for the configuration with asymmetric proton sharing that
featured the largest deviations upon microsolvation, most
VADs obtained on both grids are in very good agreement.
However the shallower region around the excess proton
of the NNP compared to the reference grid
again causes slight shifts in the VAD locations.
Since their number and also shape is still correctly reproduced
and the RDFs are in very good agreement, the
solvation of the Zundel cation by bulk helium is expected to
be described with very high accuracy by the NNP.

\section{Conclusions and Outlook}
\label{sec:cando}

In conclusion, we have presented a systematic and 
largely automated 
procedure to develop pairwise additive He-solute interaction potentials
employing high-dimensional neural networks.
This NNP approach yields very 
convincing agreement with the high level reference electronic structure method
being CCSD(T) in an essentially converged basis set
in the present cases, being the protonated water monomer and dimer. 
The flexibility of the NNs allows to easily identify deficiencies
in the training set which can be used to systematically 
improve the NNP while substantially reducing the number 
of 
expensive reference calculations.
This opens the door for fast and accurate development of
He-solute interaction potentials and therefore overcomes the obstacles of
traditional fitting approaches
used so far in this field. 
For the chosen solutes in this study, 
i.e. \hydronium{} and \zundelp{},
we were able to correctly describe the
solvation structure upon stepwise microsolvation with helium~--
from one adatom up to the solvation in bulk helium
with the same accuracy.
This enables the study of the solvation of these clusters by superfluid helium
all the way from a few helium atoms to the bulk 
superfluid. 
In addition, the presented procedure will open the door to study
other solute species solvated by quantum fluids including para-hydrogen, p-H$_2$,
and therefore will provide new insights into the nature of fundamental
inter- and intramolecular interactions in quantum solvation.

\section*{Supplementary Material}
\label{sec:si}
See the supplementary material for additional validations of the
computational methods, details on the symmetry function setup
and the comprehensive 
set
of all spatial distribution functions.

\begin{acknowledgments}
It gives us great pleasure to 
thank Harald Forbert for helpful discussions. 
This research is part of the 
Cluster of Excellence “RESOLV” (EXC 1069)
funded by the \textit{Deutsche Forschungsgemeinschaft}, DFG. 
J.B. thanks the DFG for a Heisenberg Professorship (Be3264/11-1)
and C.S. acknowledges partial financial support from
the \textit{Studienstiftung des Deutschen Volkes} as well as the
\textit{Verband der Chemischen Industrie}.
The computational resources were provided by
HPC@ZEMOS, HPC-RESOLV, BOVILAB@RUB, and RV-NRW. 
\end{acknowledgments}
\end{document}